\documentclass[aps,prd,onecolumn,nofootinbib,superscriptaddress]{revtex4}
\pdfoutput=1
\usepackage[colorlinks=true,linkcolor=blue,urlcolor=blue,filecolor=black,citecolor=red,pdfstartview=FitV,pdftitle={},pdfsubject={},pdfkeywords={},pdfpagemode=None,bookmarksopen=true]{hyperref}
\usepackage{graphicx}
\usepackage{amsmath}
\usepackage{amsfonts}
\usepackage{amssymb,ulem}
\usepackage{color}%
\usepackage{dcolumn}
\usepackage{braket}
\usepackage{eurosym}
\usepackage[usenames,dvipsnames,svgnames]{xcolor}

\newcommand{\RNum}[1]{\uppercase\expandafter{\romannumeral #1\relax}}

\usepackage[title]{appendix}
\usepackage{wrapfig,boxedminipage,setspace,subfigure,epsfig}

\begin{document}
\baselineskip=0.5 cm

\title{Distinguishing black holes with and without spontaneous scalarization in Einstein-scalar-Gauss-Bonnet theories via optical features}

\author{Xi-Jing Wang}
\email{xijingwang01@163.com}
\affiliation{Department of Astronomy, School of Physics and Technology, Wuhan University, Wuhan 430072, China}

\author{Yuan Meng}
\email{mengyuanphy@163.com}
\affiliation{Center for Gravitation and Cosmology, College of Physical Science and Technology, Yangzhou University, Yangzhou, 225009, China}

\author{Xiao-Mei Kuang}
\email{xmeikuang@yzu.edu.cn}
\thanks{corresponding author}
\affiliation{Center for Gravitation and Cosmology, College of Physical Science and Technology, Yangzhou University, Yangzhou, 225009, China}

\author{Kai Liao}
\email{liaokai@whu.edu.cn}
\thanks{corresponding author}
\affiliation{Department of Astronomy, School of Physics and Technology, Wuhan University, Wuhan 430072, China}

\begin{abstract}
Spontaneous scalarization in Einstein-scalar-Gauss-Bonnet theory admits both vacuum-general relativity (GR) and scalarized hairy black holes as valid solutions, which provides a distinctive signature of new physics in strong gravity regime. In this paper, we shall examine the optical features of Gauss-Bonnet black holes with spontaneous scalarization, which is governed by the coupling parameter $\lambda$. We find that the photon sphere, critical impact parameter and innermost stable circular orbit all decrease as the increasing of $\lambda$. Using observable data from Event Horizon Telescope, we establish the upper limit for $\lambda$. Then we construct the optical appearances of the scalarized black holes illuminated by various thin accretions. Our findings reveal that the scalarized black holes consistently exhibit smaller shadow sizes and reduced brightness compared to Schwarzschild black holes. Notably, in the case of thin spherical accretion, the shadow of the scalarized black hole is smaller, but the surrounding bright ring is more pronounced. Our results highlight the observable features of the scalarized black holes, providing a distinguishable probe from their counterpart in GR in strong gravity regime.
\end{abstract}

\maketitle

\tableofcontents
\newpage

\section{Introduction}

Remarkable detection progress on gravitational waves (GWs) \cite{LIGOScientific:2016aoc,LIGOScientific:2017vwq,LIGOScientific:2020aai} and black hole images \cite{EventHorizonTelescope:2019dse,EventHorizonTelescope:2019uob,EventHorizonTelescope:2019jan,
EventHorizonTelescope:2019ths,EventHorizonTelescope:2019pgp,EventHorizonTelescope:2019ggy,EventHorizonTelescope:2022wkp,EventHorizonTelescope:2022apq,
EventHorizonTelescope:2022wok,EventHorizonTelescope:2022exc,EventHorizonTelescope:2022urf,EventHorizonTelescope:2022xqj} means that we are entering an unprecedented era to test the physics in strong gravity regime. It has to be admitted that the effects of higher-order curvature terms are especially profound when one refers to strong field regime of gravity via detections of GWs and black hole shadows.  Such terms probably lead to the well-known ghost problem \cite{Stelle:1976gc}, but Gauss-Bonnet term is a counter-case which can be ghost-free. It contributes no dynamics to the field equations when it is minimally coupled with Einstein-Hilbert action, but considering this term to couple with a scalar field is a smart way to make it meaningful in four-dimensional spacetime. Thus, as a special scalar-tensor theory with higher derivatives \cite{Horndeski:1974wa,Deffayet:2009mn}, the Einstein-scalar-Gauss-Bonnet (EsGB) gravity involves in considerable interest. In particular, the introduction of this coupling could make the theories admit hairy compact objects such as black hole (see for examples \cite{Mignemi:1992nt,Kanti:1995vq,Torii:1996yi,Kleihaus:2011tg,Sotiriou:2013qea,Ayzenberg:2014aka,Kleihaus:2016dui}), thus, the no-hair theorem \cite{Bekenstein:1972ny,Teitelboim:1972ps} in classical general relativity (GR) becomes  more controversial.

Over the past decade, the spontaneous scalarizations with particular coupling functions in EsGB theory are extensively investigated. In this framework, besides GR solutions with a trivial scalar field configuration, the scalarized hairy solutions for black holes \cite{Doneva:2017bvd,Silva:2017uqg,Antoniou:2017acq} and other compact objects \cite{Damour:1993hw,Doneva:2017duq,Antoniou:2019awm,Ibadov:2020btp} could also exist, which indeed evade the no-hair theorem. In details, below a certain mass, the Schwarzschild black hole background may become linearly unstable in regions of strong curvature, and the scalarized branches emerge when the scalar field backreacts to the geometry \cite{Doneva:2017bvd}. The hairy branches were also found physically favorable \cite{Blazquez-Salcedo:2020rhf,Blazquez-Salcedo:2020caw,Blazquez-Salcedo:2018jnn}. To better understand the procedure of spontaneous scalarizations in EsGB theory, extensive studies have been carried on. The spontaneous scalarizations due to a coupling of a scalar field to Ricci scalar \cite{Herdeiro:2019yjy}, and to the Maxwell term \cite{Herdeiro:2018wub,Fernandes:2019rez} were studied and in the presence of Chern-Simons invariant was studied in \cite{Brihaye:2018bgc}. The spontaneous scalarization of asymptotically AdS/dS black holes with a negative/positive cosmological constant was extended in \cite{Bakopoulos:2018nui,Brihaye:2019gla,Bakopoulos:2019tvc,Bakopoulos:2020dfg,Lin:2020asf,Guo:2020sdu,Guo:2024vhq}. The influences of horizon curvature and spacetime structure on black hole spontaneous scalarization were studied in \cite{Guo:2020zqm}. Also in EsGB theory, the spin-induced black hole spontaneous scalarization, which is the outcome of linear tachyonic instability triggered by rapid rotation, was explored in \cite{Collodel:2019kkx,Dima:2020yac,Doneva:2020kfv,Herdeiro:2020wei,Berti:2020kgk,Lara:2024rwa}, and the dynamical process during the spontaneous scalarization was simulated in \cite{Liu:2022fxy}.

On the observational side, it was addressed in \cite{Wong:2022wni} that spontaneous scalarization provides a distinctive signature of new physics in the strong gravity regime for those black holes whose Schwarzschild radii is comparable to the new length scale $\lambda$ in the theory, and the theory is not easy to constrain. Even so, they used the GW events (GW190814 and GW151226) and examined the GW's constraining power on EsGB theory that allows for the spontaneous scalarization of black holes. Moreover, the shadows radius of black holes and wormholes with spontaneous scalarization in this theory have been explored in \cite{Antoniou:2022dre}, in which the authors considered four sGB couplings and provided the possible constrains or even excluded some scalarized solutions by various observational data published by the Event Horizon Telescope (EHT). However, the accretion matters surrounding around the black hole usually affect the observational appearance of the central object. To our knowledge, the optical features of the black holes with spontaneous scalarization in this theory have not been elaborated. In this regard, we are especially interested in the issue: Can the photon rings, shadows and images of the scalarized Gauss-Bonnet black holes be distinguishable from those of their counterpart in GR?

The black hole shadow corresponds to light rays that neither go to infinity nor fall into the event horizon from the view of backward ray-tracing, but are trapped within the spacetime. For the Schwarzschild black hole, its shadow shows a perfect circle \cite{Synge:1966okc}. When turning on the rotation, the shadow of Kerr black hole becomes deformed and presents $D$-shape \cite{Bardeen:1973tla}. Even though the spacetime disclosed by these shadows is in good agreement with the prediction of Kerr black hole, it still leaves some space for theoretical parameters of Kerr-like or other black hole solutions in modified theories of gravity due to the observational uncertainties. So the shadows of various modified theories of gravity have been extensively discussed, see for examples \cite{Amarilla:2010zq,Wei:2013kza,Wang:2017hjl,Dastan:2016vhb,Wang:2018prk,Kuang:2022ojj,Meng:2023wgi,Meng:2022kjs,Addazi:2021pty,
Li:2020drn,Kuang:2022xjp,Kumar:2019pjp,Shaikh:2021yux,Vagnozzi:2022moj,Sui:2023rfh,Kuang:2024ugn} and references therein.

In a realistic universe, the black holes are usually surrounded by accretion flows that determine the optical appearances of black holes. One should use general relativistic magnetohydrodynamics to simulate the complex astrophysical process between black hole and accretion flow and further extract the image of black hole \cite{EventHorizonTelescope:2019pcy}. However, some simplified accretion toy models are enough to capture main characteristics of black hole images. For example, Gralla et al. considered the Schwarzschild black hole illuminated by an optically and geometrically thin accretion disk \cite{Gralla:2019xty}. By the number ($m$) of intersections between photon and accretion disk, they classified the emission types of light rays emitted from the accretion disk: direct ($m=1$), lensed ring ($m=2$) and photon ring ($m\geq 3$) emissions. The results showed that the direct emission determines the dominant contribution to the total observed brightness, followed by lensed ring emission, and photon ring emission can almost be ignored \cite{Gralla:2019xty}. Moreover, when considering another kind of accretion, namely spherical accretion, the corresponding shadows are usually influenced by the spacetime geometry rather than the details of the accretions \cite{Falcke:1999pj,Narayan:2019imo}. So far, the observational appearances of black holes surrounded by various accretions have attracted much more attentions in modified theories of gravity \cite{Johnson:2019ljv,Zeng:2020dco,Zeng:2020vsj,Peng:2020wun,Qin:2020xzu,Chakhchi:2022fls,Guo:2021bhr,Hu:2022lek,Wen:2022hkv,Meng:2024puu,Gao:2023mjb,
Wang:2023vcv,Chen:2023qic,Yang:2024utv}. Specially and importantly, the recent studies showed that the photon ring of black hole image could produce strong and universal signatures on long interferometric baselines \cite{Johnson:2019ljv}. It is feasible to measure the structures of photon ring precisely through the analysis of interferometric signatures in black hole images.

Thus, in this paper, we will study the light rays around the scalarized Gauss-Bonnet black hole by using ray-tracing method. We will show the photon rings and images of the scalarized Gauss-Bonnet black hole illuminated by the optically and geometrically static thin accretion disk and spherical accretion, respectively. Based on the observational differences, we explore the potential method in strong field regime to distinguish the black hole with and without spontaneous scalarization by using black hole shadows and images. The remaining of this paper is organized as follows. In Sec.\ref{review}, we give a brief review on Gauss-Bonnet black hole with spontaneous scalarization in EsGB theory. In Sec.\ref{sec-Trajectories}, we calculate the radius and critical impact parameter of the photon sphere for the scalarized black hole and give the constrains on the coupling parameter by utilizing the current black hole shadows data. In Sec.\ref{thin}, we classify the light rays based on the trajectories of photons outside the scalarized black hole and investigate the optical appearances of scalarized black holes illuminated by the optically and geometrically thin accretion disk. In Sec.\ref{spherical}, we consider the scalarized black hole surrounded by the static spherical accretion and show the optical appearances. Finally, we conclude our findings in Sec.\ref{conclusion}.

\section{A quick review on scalarized Gauss-Bonnet black hole}\label{review}

We consider four dimensional EsGB theory with the action \cite{Doneva:2017bvd}
\begin{eqnarray}
&&S=\frac{1}{16\pi}\int d^4x \sqrt{-g} 
\Big[R - 2\nabla_\mu \varphi \nabla^\mu \varphi + \lambda^2 F(\varphi){\cal R}^2_{\text{GB}} \Big],\label{eq:quadratic}
\end{eqnarray}
where $R$ is the Ricci scalar, $\varphi$ is the scalar field and ${\cal R}^2_{\text{GB}}=R^2-4R^{\mu\nu}R_{\mu\nu}+R^{\mu\nu\alpha\beta}R_{\mu\nu\alpha\beta}$ is the Gauss-Bonnet invariant.
$F(\varphi)$ is the scalar coupling function which is non-minimally coupled to Gauss-Bonnet term. Thus, $\lambda$ is the Gauss-Bonnet coupling parameter that has the dimension of length. When $\lambda=0$ and $\varphi=\text{const.}$, this theory will recover to GR. By varying the action \eqref{eq:quadratic} with respect to the metric tensor $g_{\mu\nu}$ and the scalar field $\varphi$, we can derive the following fields equations \cite{Doneva:2017bvd}
\begin{eqnarray}\label{FE}
&&R_{\mu\nu}- \frac{1}{2}R g_{\mu\nu} + \Gamma_{\mu\nu}= 2\nabla_\mu\varphi\nabla_\nu\varphi -  g_{\mu\nu} \nabla_\alpha\varphi \nabla^\alpha\varphi,\\
&&\nabla_\alpha\nabla^\alpha\varphi=-\frac{\lambda^2}{4} \frac{dF(\varphi)}{d\varphi} {\cal R}^2_{\text{GB}}, \label{eqscalar}
\end{eqnarray}
where the symbol $\Gamma_{\mu\nu}$ is defined by 
\begin{eqnarray}
\Gamma_{\mu\nu}&=& - R(\nabla_\mu\Psi_{\nu} + \nabla_\nu\Psi_{\mu} ) - 4\nabla^\alpha\Psi_{\alpha}\left(R_{\mu\nu} - \frac{1}{2}R g_{\mu\nu}\right) + 
4R_{\mu\alpha}\nabla^\alpha\Psi_{\nu} + 4R_{\nu\alpha}\nabla^\alpha\Psi_{\mu} \nonumber \\ 
&& - 4 g_{\mu\nu} R^{\alpha\beta}\nabla_\alpha\Psi_{\beta} 
+ \,  4 R^{\beta}_{\;\mu\alpha\nu}\nabla^\alpha\Psi_{\beta},
\end{eqnarray}  
with $\Psi_{\mu}= \lambda^2 \frac{dF(\varphi)}{d\varphi}\nabla_\mu\varphi$. Just like in GR, the above field equations are of second-order and the theory is free from ghosts \footnote{If we replace the Gauss-Bonnet term in the action \eqref{eq:quadratic} with  another curvature term, for example, the Weyl square term defined by $C^2\equiv C^{\mu\nu\alpha\beta}C_{\mu\nu\alpha\beta}$ where $C_{\mu\nu\alpha\beta}$ is the Weyl tensor, the new theory will lead to fourth-order field equations and thus introduce ghost modes. Please see \cite{Lu:2015cqa,Wang:2023klu,Myung:2023iqc} for more details.}. It was shown that the theory will result in GR equation with standard Klein-Gordon (KG) equation and admits scalar-free GR solutions when the coupling function $F(\varphi)$ satisfies the conditions $\frac{dF}{d\varphi}\big|_{\varphi=0}=0$ and $\varphi=0$ \cite{Doneva:2017bvd}. Considering the scalar perturbation $\delta \varphi$ around the GR solution, the linearized KG equation \eqref{eqscalar} reads
\begin{equation}
\left(\square_{(0)}-m^2_{\text{eff}}\right)\delta \varphi=0,
\end{equation}
where $m^2_{\text{eff}}=-\frac{\lambda^2}{4} \frac{d^2F}{d\varphi^2}\big|_{\varphi=0} {\cal R}^2_{\text{GB}(0)}$ is effective mass squared of the scalar perturbation $\delta \varphi$ and the subscript ``0" represents the GR background geometry. It was shown that the negative effective mass triggers a tachyonic instability, the scalar-free GR solution is unstable and the BH solutions with nontrivial scalar field will be produced spontaneously \cite{Doneva:2017bvd}. That is the spontaneous scalarization of BH induced by the curvature of the spacetime \cite{Doneva:2017bvd,Silva:2017uqg} we will focus on, which is different from that sourced by matters \cite{Cardoso:2013fwa,Cardoso:2013opa}.

Next we will closely follow the steps described in \cite{Doneva:2017bvd} to construct the scalarized black hole solutions. We consider the following ansatz to construct the static, spherically symmetric and asymptotically flat spacetime and assume that the scalar field has the same symmetric configurations with the spacetime given by
\begin{equation}
ds^2=-h(r)dt^2+\frac{1}{f(r)}dr^2+r^2\left(d\theta^2+\text{sin}^2\theta d\phi^2\right),  \qquad \varphi=\varphi(r). \label{ansatz}
\end{equation}
The asymptotic flatness requires the metric functions and scalar field to behave as
\begin{equation}
h(r)=1-\frac{2M}{r}+\cdots,\quad
f(r)=1-\frac{2M}{r}+\cdots,\quad
\varphi(r)=\varphi_{\infty}+\frac{D}{r}+\cdots, \label{expanInf}
\end{equation}
where $M$ and $D$ are the mass and scalar charge, respectively, and the constant of scalar field at infinity, $\varphi_\infty$, can be chosen as zero. Moreover, the existence of event horizon requires $h(r)|_{r \rightarrow r_h} \rightarrow 0$ and $f(r)|_{r \rightarrow r_h} \rightarrow 0$, while the regularity of scalar field near the event horizon directly gives
\begin{eqnarray}
\left(\frac{d\varphi}{dr}\right)_{h}= \frac{r_{h}}{4 \lambda^2 \frac{dF}{d\varphi}(\varphi_{h})} 
\left[-1 \pm \sqrt{1 - \frac{24\lambda^4}{r^4_{h}} \left(\frac{dF}{d\varphi}(\varphi_{h})\right)^2}\right],
\end{eqnarray} 
where the subscript $h$ indicates the physical quantities evaluated at the event horizon.
Note that one has to choose the plus sign, since
only in this case can we recover the Schwarzschild solution
in the limit $\varphi_h \rightarrow 0$. Meanwhile, the above equation implies that the black hole with a nontrivial scalar field can exist only when
\begin{equation}
r_h^4 > 24 \lambda^4 \left(\frac{dF}{d\varphi}(\varphi_{h})\right)^2. \label{eq-regbcs}
\end{equation}

\begin{figure}[htbp]
\centering
{\includegraphics[width=6cm]{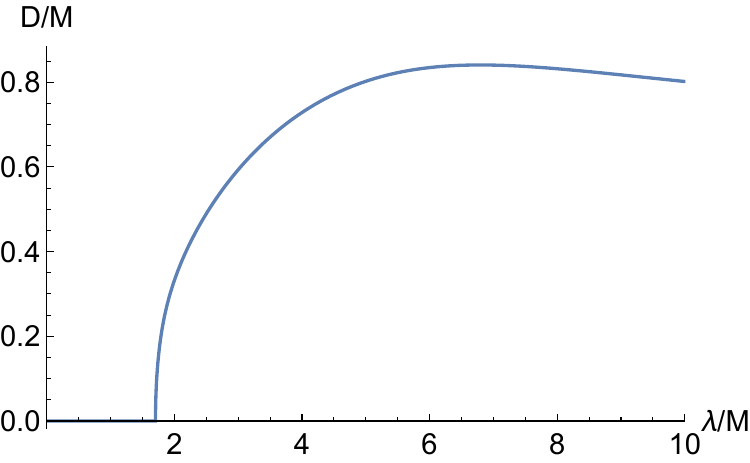}}\hspace{5mm}
{\includegraphics[width=6cm]{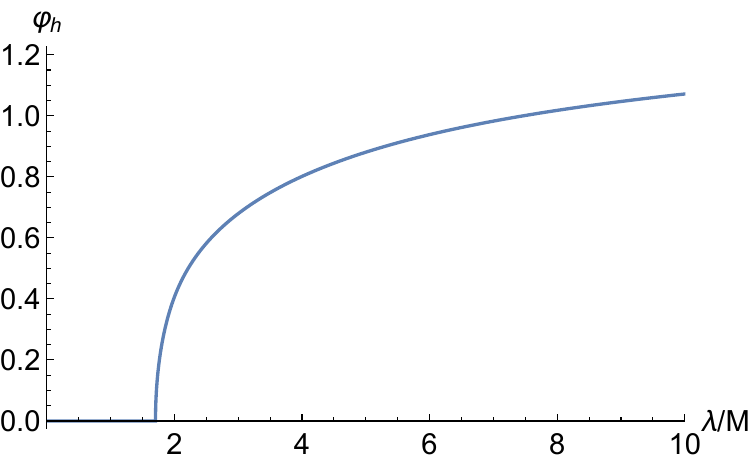}}
\caption{The scalar charge $D$ and the scalar field at the event horizon $\varphi_h$ as the functions of coupling $\lambda$.}
\label{figcharge}
\end{figure}

\begin{figure}[htbp]
\centering
{\includegraphics[width=4.5cm]{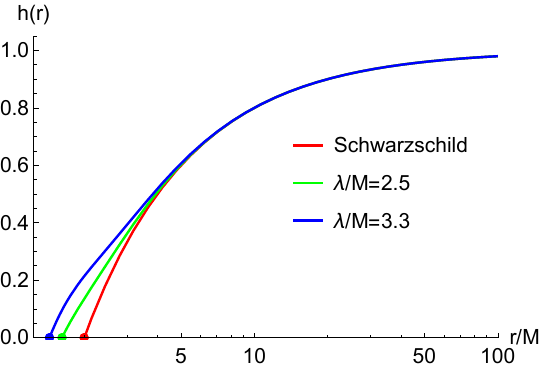}}\hspace{5mm}
{\includegraphics[width=4.5cm]{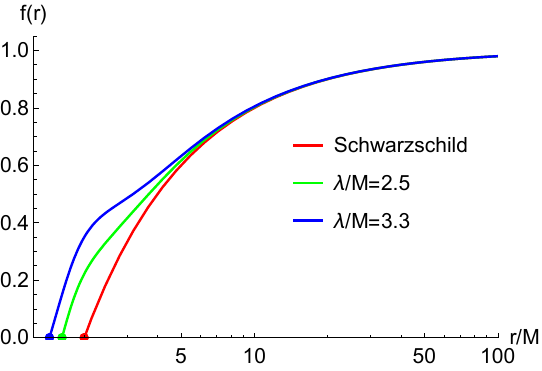}}\hspace{5mm}
{\includegraphics[width=4.5cm]{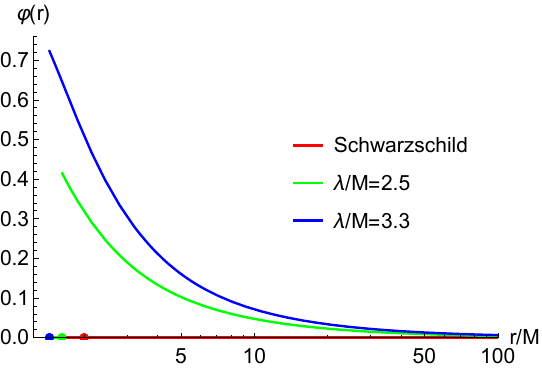}}
\caption{The metric and scalar fields functions as the functions of radial coordinate $r$ for different coupling $\lambda$. The color dots denote the positions of event horizon of corresponding coupling parameters.}
\label{metricscalar}
\end{figure}

Thus, one can solve the field equations with the aforementioned boundary conditions to find  the scalarized black hole solutions, once  the formula of coupling function $F(\varphi)$ is given. A quite similar choice 
in the spontaneous scalarization of neutron stars \cite{Damour:1993hw}
\begin{equation}\label{eq:coupling function}
F(\varphi)=\frac{1}{12}\left(1-e^{-6 \varphi^2}\right),
\end{equation}
was found to lead the significant deviations from the Schwarzschild solution and the condition \eqref{eq-regbcs} is satisfied for a large enough range of parameters \cite{Doneva:2017bvd}. It is noted that in the following study, we will rescale the quantities $(r, \lambda, D, M)$ of this system into the dimensionless ones $(r/M, \lambda/M, D/M, 1)$. The numerical results of  the scalar charge $D$ and scalar field at the event horizon $\varphi_h$  as functions of the coupling parameter are plotted in Fig.\ref{figcharge} \footnote{The $\mathcal{Z}_2$ symmetry, i.e., the invariance under the transformation $\varphi \rightarrow -\varphi$, of  this theory allows us to focus on the positive branches of $D$ and $\varphi_h$.}. From this figure, one can find that when the coupling $\lambda$ is small, the scalar field is trivial, implying the Schwarzschild black hole solution. When the coupling is larger than a critical value ($\lambda/M\approx 1.7$ agrees well with that found in \cite{Doneva:2017bvd}), nontrivial scalar field emerges, indicating that the scalarized black holes exist due to the spontaneous scalarization of the Schwarzschild black hole. Samples of profiles of the scalarized black hole metric and the nontrivial scalar field are depicted in Fig.\ref{metricscalar} where we also exhibit the Schwarzschild case for comparison. It is obvious that the scalarized black holes  deviate significantly from the Schwarzschild solution, but they share the same flat asymptotic behavior.

Notably, Fig.\ref{metricscalar} only gives the first nontrivial branch of the scalarized solutions in which the scalar field has solo zero in the infinity, but many other nontrivial branches can exist in certain region of the parameter, depending on the number of zeros of the scalar field. However, the first nontrivial branch was found to be thermodynamically favorable over both their counterpart in GR and other branches, of which the metric profiles are also indistinguishable from Schwarzschild solution \cite{Doneva:2017bvd}. Therefore, we will only focus on the first nontrivial branch of scalarized solutions to explore the effect of spontaneous scalarization on the optical features of black hole.

\section{Photon sphere of the scalarized Gauss-Bonnet black hole}\label{sec-Trajectories}

In this section, we will investigate the radius and critical impact parameter of photon sphere for the scalarized Gauss-Bonnet black hole. The motions of photons are determined by the Euler-Lagrange equation
\begin{equation}\label{eq-ELeq}
\frac{d}{d\tau}\left(\frac{\partial \mathcal{L}}{\partial \dot{x}^\mu}\right)=\frac{\partial \mathcal{L}}{\partial x^\mu},
\end{equation}
where $\tau$ is the affine parameter and $\dot{x}^\mu=d{x}^\mu/d\tau$ denotes the four-velocity of the photon. For the metric \eqref{ansatz}, the Lagrangian of photons takes the following form
\begin{equation}
\mathcal{L}=\frac{1}{2}g_{\mu\nu}\dot{x}^\mu\dot{x}^\nu=\frac{1}{2}\left(-h(r)\dot{t}^2+\frac{1}{f(r)}\dot{r}^2+r^2\left(\dot{\theta}^2+\text{sin}^2\theta \dot{\phi}^2\right)\right).\label{lag}
\end{equation}
Considering the time-translational and spherical symmetries of the spacetime, there exist two conserved quantities of photons which are the energy $E$ and angular momentum $L_z$ respectively given by
\begin{equation}
E\equiv -\frac{\partial \mathcal{L}}{\partial \dot{t}}=h(r)\dot{t},   \qquad  L_z\equiv \frac{\partial \mathcal{L}}{\partial \dot{\phi}}=r^2\text{sin}^2\theta\dot{\phi}.\label{conserved}
\end{equation}
Moreover, due to the spherical symmetries of the metric, it is convenient to only focus on the photons moving on the equatorial plane ($\theta=\pi/2$). Further, from the $\mathcal{L}=0$ for photons and impact parameter $b$ defined by $b\equiv L_z/E$, we can finally obtain three first-order differential equations that determine the photons' motion in the spacetime written by
\begin{align}
&\dot{t}=\frac{1}{b\, h(r)},\label{eq1}\\
&\dot{\phi}=\pm\frac{1}{r^2},\label{eq2}\\
&\frac{h(r)}{f(r)}\dot{r}^2+V_{\text{eff}}(r)=\frac{1}{b^2},\label{eq3}
\end{align}
with the effective potential
\begin{equation}
V_{\text{eff}}(r)=\frac{h(r)}{r^2}.\label{eqveff}
\end{equation}
Note that the affine parameter $\tau$ has been redefined as $\tau/L_z$. The sign $``+"$ and $``-"$ denote that the photon moves on the equatorial plane along the counterclockwise and clockwise direction, respectively. From the Eq.\eqref{eq2} and Eq.\eqref{eq3}, we can obtain the compact form for the equation of motion given by
\begin{equation}
\frac{dr}{d\phi}=\pm r^2\sqrt{\frac{f(r)}{h(r)}\left(\frac{1}{b^2}-V_{\text{eff}}(r)\right)}.\label{eqphoton}
\end{equation}
Obviously, the trajectory of photon is determined by the impact parameter $b$ and effective potential $V_{\text{eff}}$. When $dr/d\phi=0$, meaning $V_{\text{eff}}(r_0)=1/b^2$. It implies that the photon approaches the black hole from infinity, and then moves towards infinity again after their distance reaching a minimum radius $r_0$. Further, there exists a critical value of $r_0$, smaller than which will make the photon fall into the black hole instead of escaping to infinity. This critical value is known as radius of photon sphere $r_{ph}$ that corresponds to the critical impact parameter $b_{ph}$, which is determined by $d^2r/d\phi^2=0$ combined with $dr/d\phi=0$. These two conditions of photon sphere will be finally translated to
\begin{equation}
V_{\text{eff}}(r_{ph})=\frac{1}{b_{ph}^2}, \qquad V_{\text{eff}}'(r_{ph})=0.  \label{formulabrph}
\end{equation}
While for the massive particle, similar physical process gives the innermost stable circular orbit (ISCO), the radius of which for the particle orbiting the metric \eqref{ansatz} can be evaluated by \cite{Wang:2023vcv}
\begin{equation}
r_{isco}=\frac{3h(r_{isco})h'(r_{isco})}{2h'(r_{isco})^2-h(r_{isco})h''(r_{isco})}.\label{formularisco}
\end{equation}

\begin{figure}[htbp]
\centering
\subfigure[\, $r_h$]
{\includegraphics[width=3.8cm]{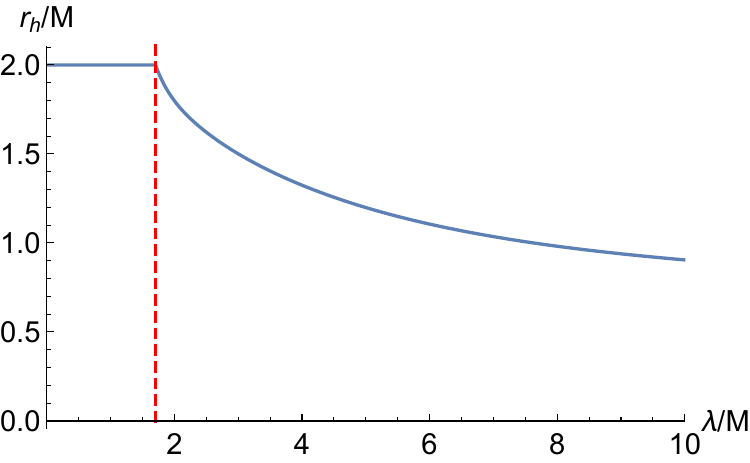}}\hspace{3mm}
\subfigure[\, $r_{ph}$] 
{\includegraphics[width=3.8cm]{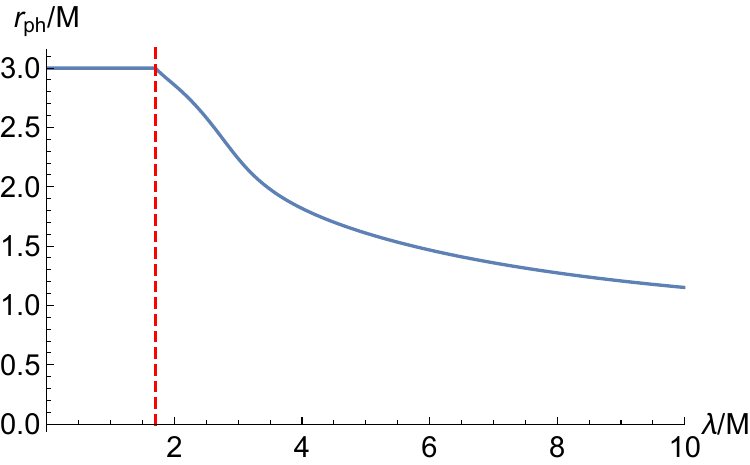}}\hspace{3mm}
\subfigure[\, $b_{ph}$] 
{\includegraphics[width=3.8cm]{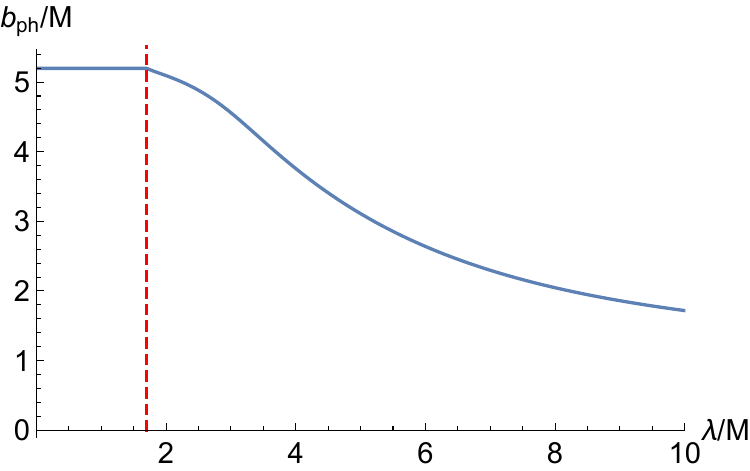}}\hspace{3mm}
\subfigure[\, $r_{isco}$] 
{\includegraphics[width=3.8cm]{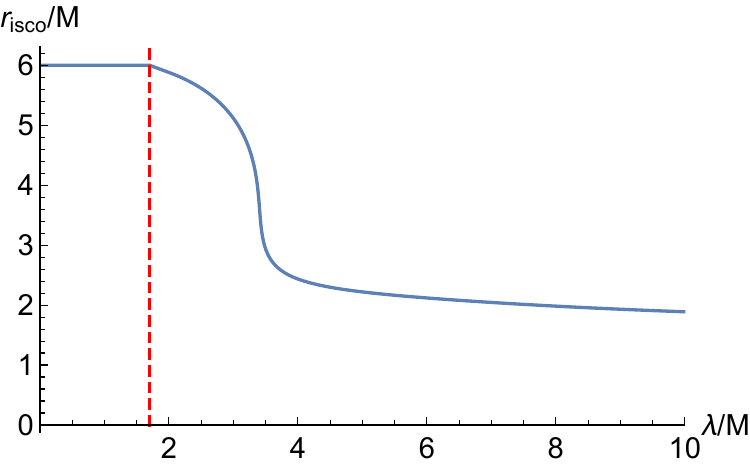}}
\caption{The event horizon $r_h$, radius of photon sphere $r_{ph}$, critical impact parameter $b_{ph}$, and innermost stable circular orbit $r_{isco}$ as the functions of the coupling parameter $\lambda$. The vertical red dashed lines denote the critical coupling for the formation of spontaneous scalarization $\lambda/M=1.7$.}
\label{figquantity2}
\end{figure}

The radii of event horizon $r_h$, photon sphere $r_{ph}$, critical impact parameter $b_{ph}$, and ISCO $r_{isco}$ as the functions of coupling parameter $\lambda$ are shown in Fig.\ref{figquantity2}. Samples of numerical results are listed in Table \ref{BHquantity}. We find that when the coupling parameter is increased from zero, these four physical quantities keep unchanged since the Schwarzschild black hole always keeps its own state in certain parameter region. When the increased coupling parameter is beyond critical coupling for the formation of spontaneous scalarization, these four physical quantities decrease monotonously.

\begin{table}[htbp]
\begin{center}
\begin{tabular}{|c|c|c|c|c|c|c|c|c|c|c|}
\hline
{$\lambda/M$}&{Schwarzschild}   &{$2$}   &  {$2.5$}   &  {$3$}   &  {$3.3$} &  {$4$} \\
\hline
{$r_{h}/M$}  &  {2}   &{1.79415}   &{1.62264}   &{1.50074}    &{1.44096} &{1.32459} \\
\hline
{$r_{ph}/M$} &  {3}   &{2.8531}   &{2.58255}   &{2.23677}    &{2.06705} &{1.81719}\\
\hline
{$b_{ph}/M$}  &  {5.19615}   &{5.08947}   &{4.88294}   &{4.55848}    &{4.31981} &{3.75795}\\
\hline
{$r_{isco}/M$}  &  {6}   &{5.87473}   &{5.61554}   &{5.11077}    &{4.41099} &{2.43372}\\

\hline
\end{tabular} \\
\caption{The numerical results of the event horizon $r_h$, radius of photon sphere $r_{ph}$, critical impact parameter $b_{ph}$ and innermost stable circular orbit $r_{isco}$ as the functions of the coupling parameter $\lambda$.}\label{BHquantity}
\end{center}
\end{table}

\begin{figure}[htbp]
\centering
{\includegraphics[width=6cm]{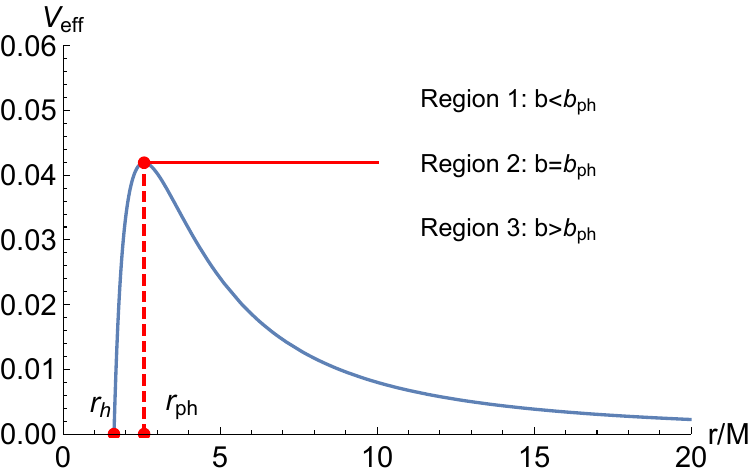}}\hspace{5mm}
{\includegraphics[width=6cm]{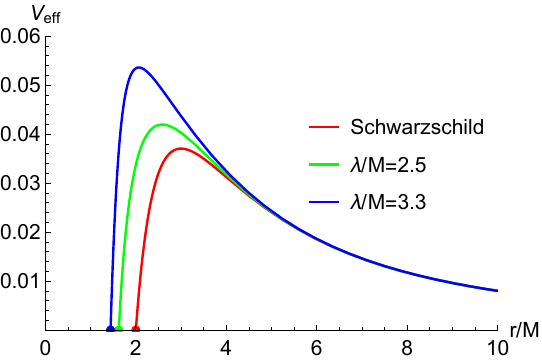}}
\caption{\textbf{Left:} The effective potential $V_{\text{eff}}$ as a function of radial coordinate $r$ for $\lambda/M=2.5$ as an example. \textbf{Right:} The effective potential as a function of radial coordinate $r$ for different $\lambda$, where the color dots denote the positions of event horizon of corresponding coupling parameters.}
\label{figVeff}
\end{figure}

We analyze the motion of photon from the point of the effective potential as shown in the left figure of Fig.\ref{figVeff}. In Region 3 ($b>b_{ph}$), the photon from the infinity will be scattered into infinity after passing through the turning point. In Region 1 ($b<b_{ph}$), the photon will fall into the black hole. However for the critical situation in Region 2 ($b=b_{ph}$), the photon asymptotically approaches the photon sphere and then orbits around the black hole infinite times. These analyses are in good agreement with the trajectories of light rays by numerically solving Eq.\eqref{eqphoton}. In the right figure of Fig.\ref{figVeff}, we show the effective potential for different coupling $\lambda$ and find that when increasing $\lambda$, the position of the effective potential's peak moves left and the peak is improved, which implies the decreasing of $r_{h}$, $r_{ph}$ and $b_{ph}$ shown in Fig.\ref{figquantity2}. Note that we scan the coupling parameter space of scalarized Gauss-Bonnet black hole and find that the effective potential always has a single-peak that indicates the existence of only one photon sphere. This is different from the scalarized Reissner-Nordstr\"{o}m (RN) black hole \cite{Gan:2021pwu,Gan:2021xdl} and the hairy Schwarzschild black hole \cite{Meng:2023htc} that will show the double-peak in the effective potential and lead to the structure of multiple photon spheres and more richer optical features.

\begin{figure}[htbp]
\centering
{\includegraphics[width=6cm]{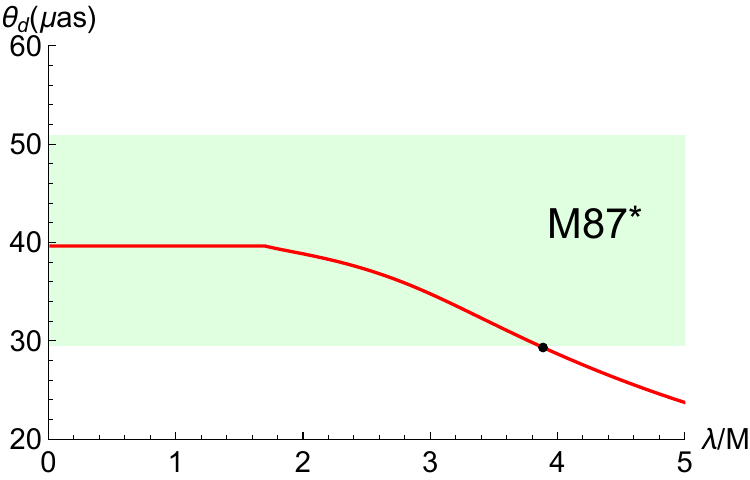}}\hspace{5mm}
{\includegraphics[width=6cm]{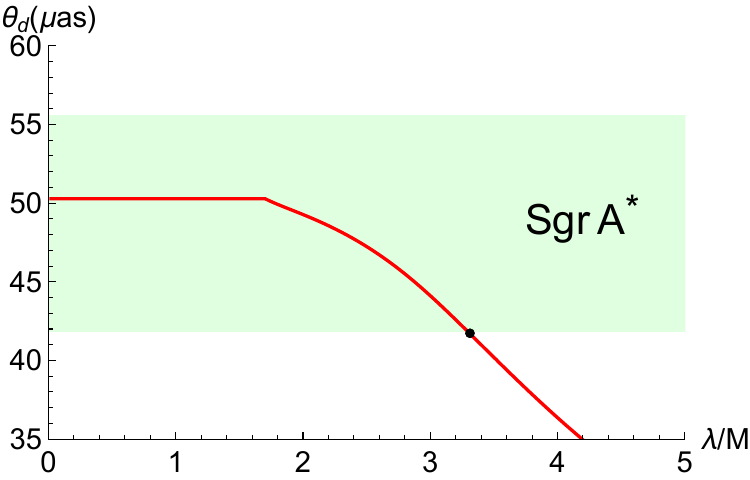}}
\caption{The constraints of the coupling parameter $\lambda$  by the observational shadows of M87* (\textbf{Left}) and Sgr A* black holes (\textbf{Right}). The black dots denote the upper limits for $\lambda$, which are $\lambda/M=3.89$ and $\lambda/M=3.31$, respectively.}
\label{figEHT}
\end{figure}

For a distant observer, the angular diameter $\theta_d$ of black hole shadow is measured by \cite{Kumar:2020owy}
\begin{equation}
\theta_d=2\frac{b_{ph}}{d_L}, \label{eqangular}
\end{equation}
where $b_{ph}$ is the critical impact parameter and $d_L$ is the distance between the black hole and observer. We can use the observational data on the black hole shadow from the EHT to roughly constrain the coupling parameter $\lambda$. Though in \cite{Antoniou:2022dre}, various bounds of shadow radius measured by the black hole mass, due to different  instruments or algorithms for the image processing, especially for Sgr A* black hole were summarized, here we shall use the  published angular diameters directly evaluated from \eqref{eqangular}, which are also intuitive. For the M87* black hole with the mass $M=6.5\times 10^9 M_\odot$ and distance $d_L=16.8\,\text{Mpc}$ \cite{EventHorizonTelescope:2019dse}, the angular diameter is roughly bounded between $29.32\,\mu as$ and $51.06\,\mu as$ \cite{EventHorizonTelescope:2021dqv,Kuang:2022ojj}. While for the Sgr A* black hole with the mass $M=4.0\times 10^6 M_\odot$ and distance $d_L=8.15\,\text{kpc}$ \cite{EventHorizonTelescope:2022wkp}, the angular diameter is $\theta_d=48.7\pm 7 \,\mu as$ \cite{EventHorizonTelescope:2022wkp,EventHorizonTelescope:2022xqj}. Thus, the EHT data provides the constraints for the coupling parameter, and gives the upper limits $\lambda/M=3.89$ by M87* black hole and $\lambda/M=3.31$ by Sgr A* black hole, respectively (see Fig.\ref{figEHT}). These constraints are important because in our following studies,  we will choose the coupling parameters which  satisfy both of these EHT bounds, i.e., fall in both green areas in the plots. It is noted that the coupling function $F(\varphi)$ is also significantly relevant, since as addressed in \cite{Antoniou:2022dre} that the scalarized solutions due to the linear coupling, quadratic coupling, dilatonic coupling and quartic coupling all may be completely excluded by certain bounds of the observational shadow radius measured by the black hole mass.

\section{Rings and images of scalarized Gauss-Bonnet black hole illuminated by thin disk accretions}\label{thin}

In this section, we shall discuss the optical appearances of the scalarized Gauss-Bonnet black holes illuminated by the optically and geometrically thin accretion disks. We assume that the disk is located on the equatorial plane around the  black hole and keeps rest, viewed face-on. To this end, based on the times that photons intersects with the accretion disk, we will firstly classify the light rays, which should contribute differently to the total observed intensity, and then explore the image of the scalarized black hole surrounded by three toy accretion disks.

\subsection{Classification of light rays: direct, lensed ring and photon ring emissions}

For the certain impact parameter $b$ of the photon, one can obtain the complete  trajectory of light ray by numerically solving Eq.\eqref{eqphoton}, and also calculate the total change of azimuthal angle $\phi$ that the photon orbits the black hole. By defining the orbit number $n=\phi/(2\pi)$ proposed in \cite{Gralla:2019xty}, we can classify the light rays into three types. The first type is called direct emission with $n<3/4$, where light ray intersects the accretion disk at most once. The second type with $3/4<n<5/4$ is lensed ring emission, where light ray crosses the accretion disk twice. The third type is photon ring emission where the light ray with $n>5/4$ crosses the accretion disk at least three times. Please refer to \cite{Hu:2022lek,Wielgus:2021peu,Bisnovatyi-Kogan:2022ujt} for the schematic diagrams to help understand above descriptions.

\begin{figure}[htbp]
\centering
{\includegraphics[width=5cm]{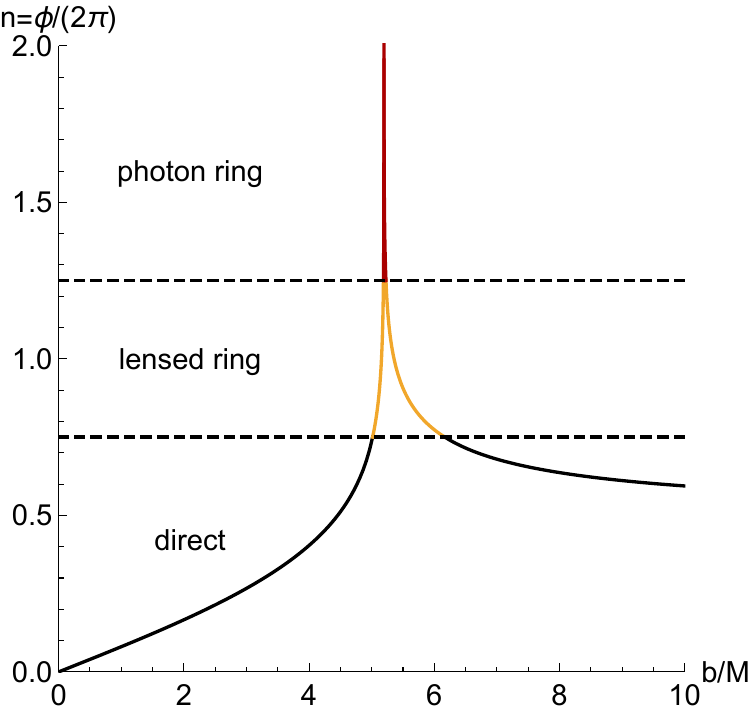}\hspace{3mm}  \includegraphics[width=5cm]{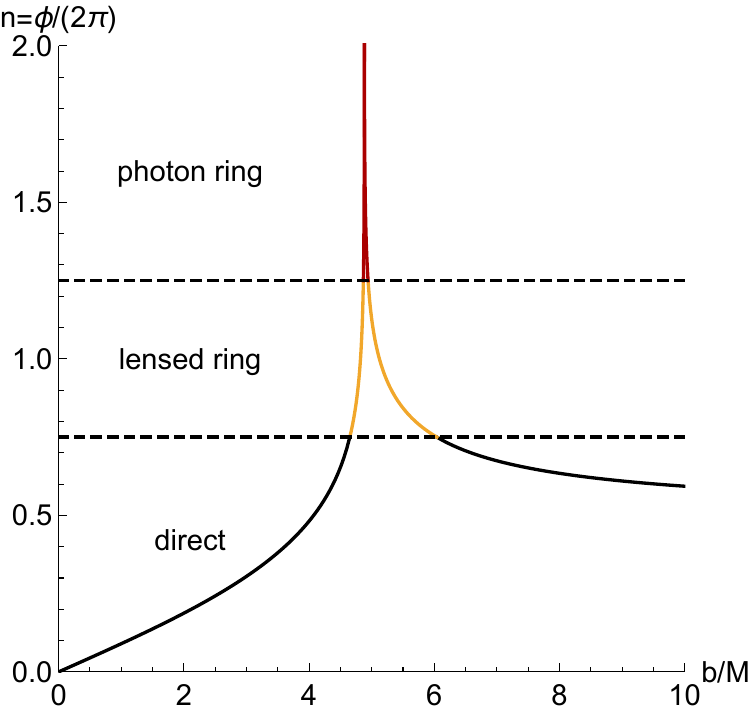}\hspace{3mm}
\includegraphics[width=5cm]{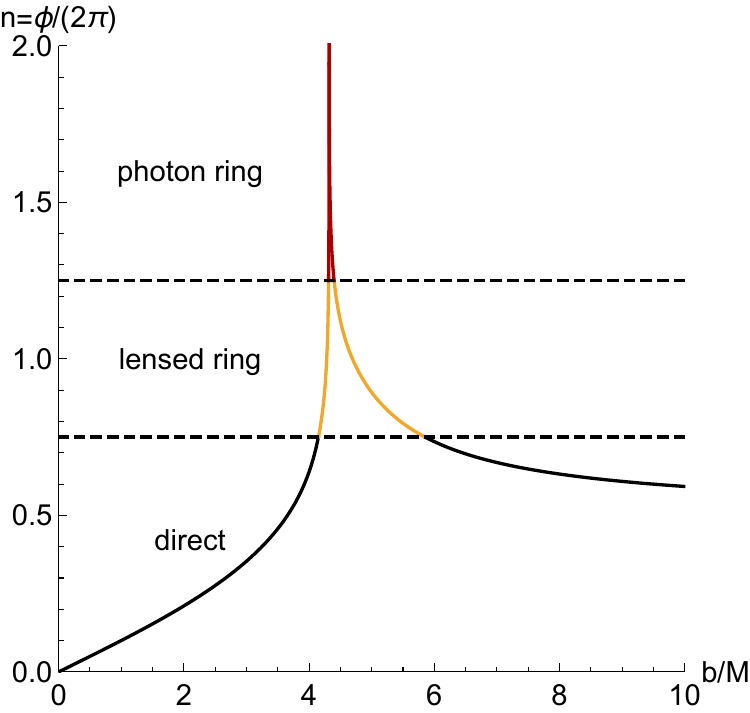}}\\
\subfigure[\, Schwarzschild]
{\includegraphics[width=5cm]{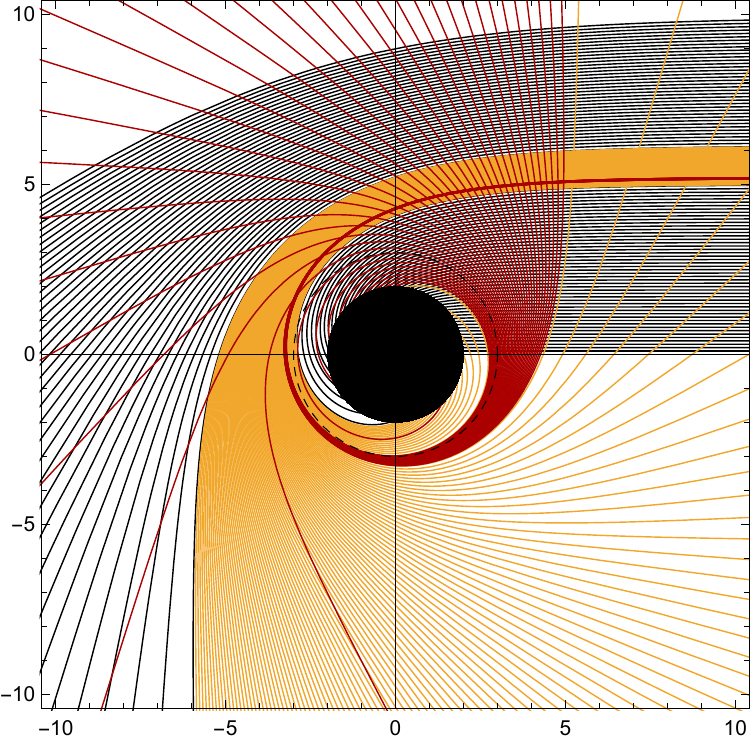}}\hspace{3mm}
\subfigure[\, $\lambda/M=2.5$]
{\includegraphics[width=5cm]{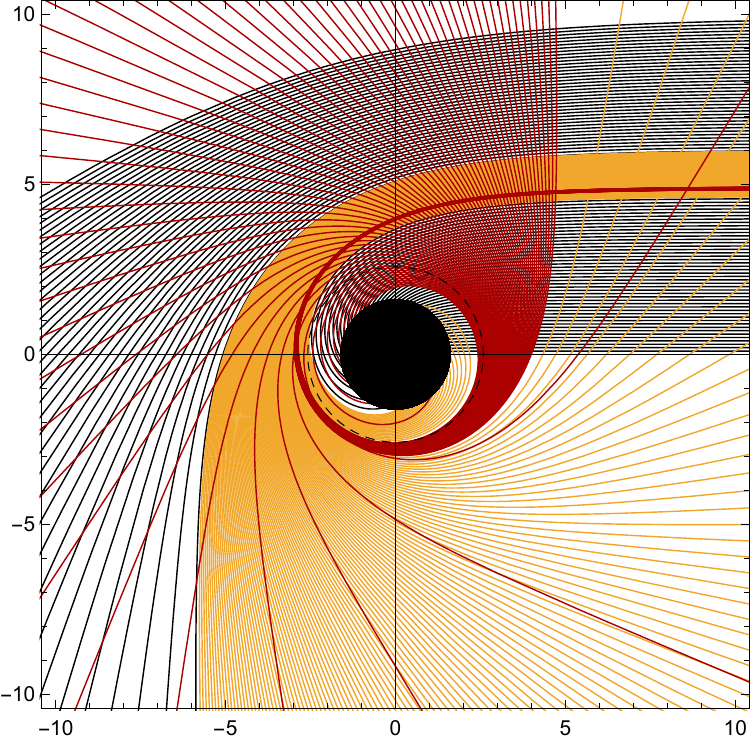}}\hspace{3mm}
\subfigure[\, $\lambda/M=3.3$]
{\includegraphics[width=5cm]{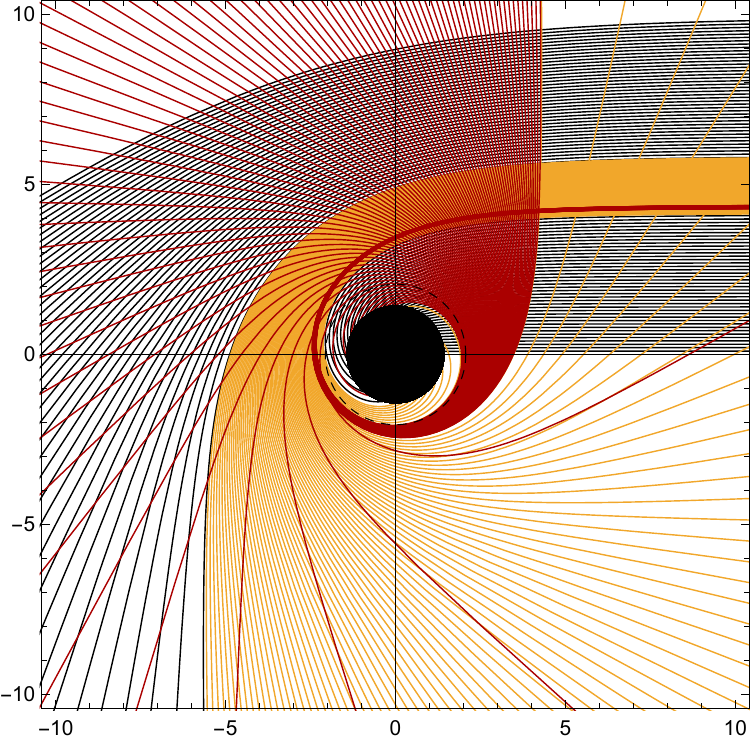}}
\caption{The orbit numbers $n$ (\textbf{Top}) and trajectories of photons (\textbf{Bottom}) as the functions of impact parameter $b$ for different coupling $\lambda$. The black, gold, and red curves correspond to the direct ($n<3/4$), lensed ring ($3/4<n<5/4$), and photon ring ($n>5/4$) emissions respectively. The black hole is shown as the solid black disk and the photon sphere is denoted by a black dashed curve.}
    \label{orbitNo}
\end{figure}

The orbit numbers with respect to impact parameter $b$ for different coupling parameter $\lambda$ are shown in the top panel of Fig.\ref{orbitNo}. Similar to that in Schwarzschild case, the scalarized Gauss-Bonnet black holes have a single-peak orbit number due to the single-peak effective potential indicating the single photon sphere. 
When increasing $\lambda$, the position of the peak moves towards the origin that indicates smaller photon sphere. Compared to the Schwarzschild black hole, the scalarized Gauss-Bonnet black holes have the wider ranges of impact parameter for both photon and lensed rings emissions, while the direct emission shrinks, which are also listed in Table.\ref{tableb}. Moreover, we also present the photon trajectories in the polar coordinates in the bottom panel of Fig.\ref{orbitNo}, which further verifies our above discussions.

\begin{table}[htbp]
\begin{center}
\begin{tabular}{|c|c|c|c|c|}
\hline
{$\lambda/M$}   &{Direct ($n<3/4$)}   &  {Lensed ring ($3/4<n<5/4$)}   &  {Photon ring ($n>5/4$)}\\
\hline
{Schwarzschild}   &{$b<5.01514$ and $b>6.16757$}   &  {$5.01514<b<5.18781$ and $5.22794<b<6.16757$}   &  {$5.18781<b<5.22794$} \\
\hline
{$2.5$}  &{$b<4.65037$ and $b>6.03995$}   &  {$4.65037<b<4.86696$ and $4.94237<b<6.03995$}   &  {$4.86696<b<4.94237$} \\
\hline
{$3.3$}  &{$b<4.14904$ and $b>5.85411$}   &  {$4.14904<b<4.31176$ and $4.40013<b<5.85411$}   & {$4.31176<b<4.40013$} \\
\hline
\end{tabular} \\
\caption{The ranges of impact parameter $b$ correspond to the direct, lensed ring, and photon ring emissions of the scalarized black hole for different coupling $\lambda$.}\label{tableb}
    \end{center}
\end{table}

\subsection{Observed intensities and optical appearances }

In the previous subsection, we have classified the light rays in terms of the times that photons intersects with the thin accretion disk. During the process of intersection, the photon will extract energy from the accretion disk each time \cite{Gralla:2019xty}. Therefore different types of light rays will contribute differently to the total observed intensity. Moreover, the analysis in above subsection implies that compared to the Schwarzschild case, the spontaneous scalarization of black hole has a significant effect on the distribution of light rays. So in this section, we will further study the observed intensities and optical appearances of the scalarized Gauss-Bonnet black hole.

\begin{figure}[htbp]
\centering
\subfigure[\, Schwarzschild]
{\includegraphics[width=5cm]{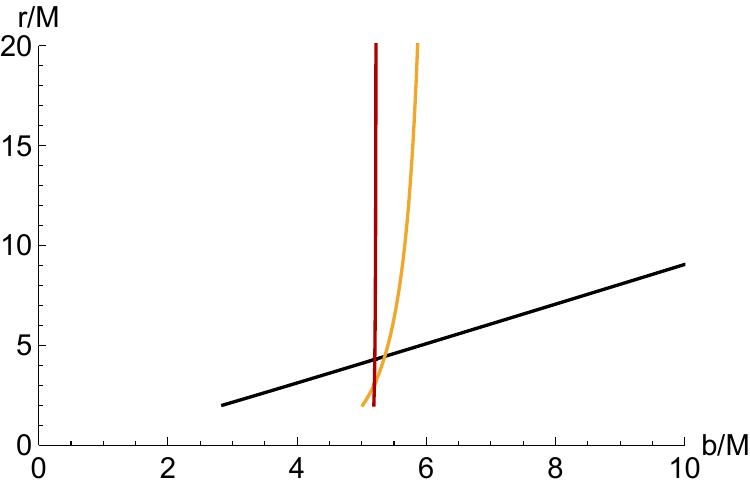}}\hspace{5mm}
\subfigure[\, $\lambda/M=2.5$]
{\includegraphics[width=5cm]{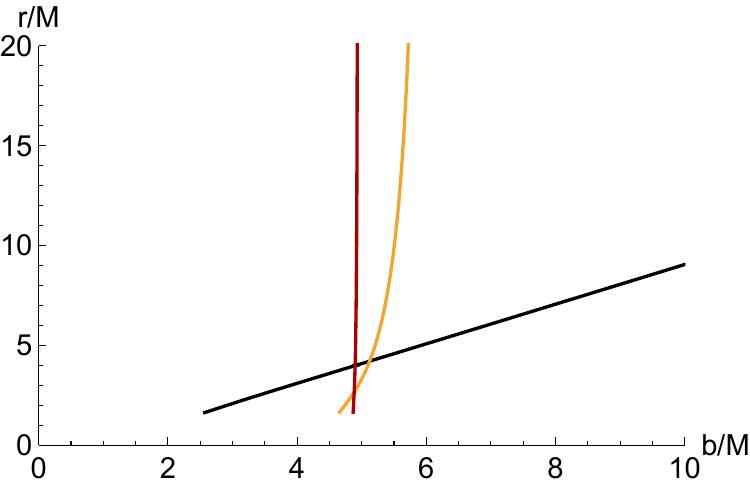}}\hspace{5mm}
\subfigure[\, $\lambda/M=3.3$]
{\includegraphics[width=5cm]{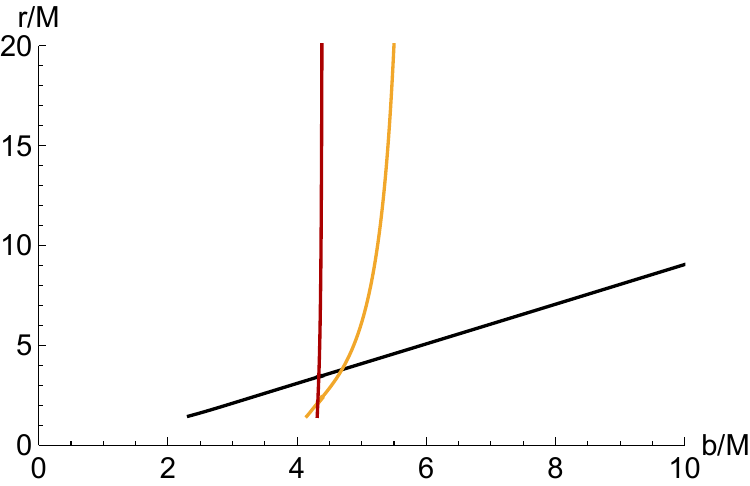}}\hspace{5mm}
\caption{The first three transfer functions in scalarized Gauss-Bonnet black hole for different $\lambda$. They represent the radial coordinate of the first (black), second (gold), and third (red) intersections with the emission disk.}
\label{figtransfer}
\end{figure}

For simplicity, we assume that the accretion disk is optically and geometrically thin and emits isotropically in the rest frame of static worldlines. The specific intensity received by the observer with emission frequency $\nu_e$ is
\begin{equation}
    I_{o}(r, \nu_o)=g^3 I_{e}(r,\nu_e),
\end{equation}
where $g=\nu_o/\nu_e=\sqrt{h(r)}$ is the redshift factor \cite{Gan:2021pwu}, and $I_{e}(r,\nu_e)$ is the specific intensity of the accretion disk. By integrating all observed frequencies of $I_{o}(r, \nu_o)$, the total observed intensity can be written as
\begin{equation}
I_{obs}(r)=\int I_{o}(r, \nu_o) d\nu_o=\int g^4 I_{e}(r,\nu_e) d\nu_e=h(r)^2 I_{em}(r),
\end{equation}
where we denote $I_{em}(r)=\int I_{e}(r,\nu_e) d\nu_e$ as the total emitted intensity, which is dependence of concrete emission profile. Based on our previous discussion, the photons that extract brightness once, twice or more to contribute to the total observed intensity depend on their impact parameter $b$. Therefore, the total observed intensity should be the sum of the intensities from each intersection read as 
\begin{equation}
I_{obs}(b)=\sum_{m}h(r)^2I_{em}(r)|_{r=r_m (b)}, \label{eqtransfer}
\end{equation}
where we have introduced the transfer function $r_m(b)$ $(m=1, 2, 3, ...)$. Physically, the transfer function, which transfers the brightness from accretion disk to the observer, establishes a one-to-one correspondence between the light ray's impact parameter $b$ and the radial coordinate $r$ of the $m$-th intersection with the accretion disk. That is to say, if giving a impact parameter, one can directly obtain the brightness after using the transfer function by Eq.\eqref{eqtransfer}. We plot the first three transfer functions $r_m (b)$ for different coupling $\lambda$ in Fig.\ref{figtransfer}. From these figures, we can find the following characteristics. For the first transfer function $(m=1)$, it corresponds to the direct image originating from direct, lensed and photon rings emissions. And its slope $dr/db$ that describes the demagnification factor at each $b$ \cite{Gralla:2019xty} is almost 1 so this direct image is the source profile after redshift. For the second transfer function $(m=2)$, it originates from lensed and photon rings emissions and its slope is large. For the third transfer function $(m=3)$, it only originates from photon ring emission and has the extremely large slope. These properties
mean that the first transfer function will play a leading contribution in the total brightness and other transfer functions contribute very little fluxes. Moreover, for the scalarized black hole, we find that the demagnification factors of the second and third transfer functions are suppressed by $\lambda$. It will lead to the wider range of the corresponding emission (that is impact parameter) than the Schwarzschild case, which agrees with the information from Table.\ref{tableb}, and make them easier to be seen. Note that due to the extreme demagnification for higher-order transfer function $(m>3)$, it is enough for our illustrative purpose to consider the contributions of the first three transfer functions into total observed intensity.

\begin{figure}[htbp]
    \centering
    \includegraphics[width=6.5cm]{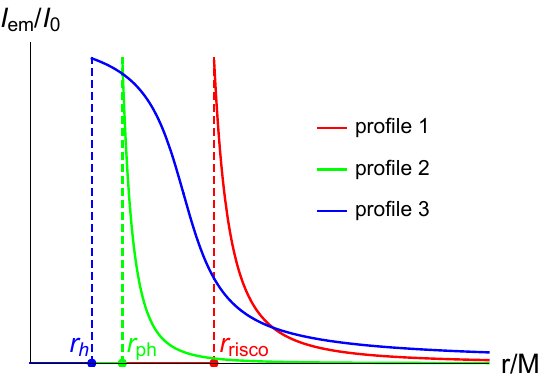}
    \caption{The sketches of three accretion disk emission profiles plotted by Eq.\eqref{diskprofile1}, Eq.\eqref{diskprofile2}, and Eq.\eqref{diskprofile3} respectively.}
    \label{figprofile}
\end{figure}

Before figuring out the optical appearance of scalarized black hole, we need provide the emission function of the accretion disk $I_{em}(r)$ appeared in Eq.\eqref{eqtransfer}. Here we consider three toy-model emission functions \cite{Wang:2022yvi,Yang:2022btw}. Firstly, we assume that the emission of accretion disk is the square decay function that starts from the innermost stable circular orbit $r_{isco}$
\begin{align}
    \text{profile 1:}\quad I_{em}(r)&:=
    \begin{cases}
        I_0\left[\frac{1}{r-(r_{ isco}-1)}\right]^2, &\hspace{1.1cm}  r>r_{ isco}\\
        0,&\hspace{1.1cm} r \leqslant r_{ isco}
    \end{cases},\label{diskprofile1}
\end{align}
where $I_0$ is the maximum intensity (the same below). Secondly, the emission function starts from the photon sphere $r_{ph}$ and exhibit a cubic decay behaviour
\begin{align}
    \text{profile 2:}\quad I_{ em}(r)&:=
    \begin{cases}
        I_0\left[\frac{1}{r-(r_{ ph}-1)}\right]^3, &\hspace{1.15cm}  r>r_{ ph}\\
        0,&\hspace{1.15cm} r\leqslant r_{\rm ph}
    \end{cases}.\label{diskprofile2}
\end{align}
Thirdly, the emission function show more moderate decay than above two functions emitting from the event horizon $r_h$
\begin{align}
    \text{profile 3:}\quad I_{ em}(r)&:=
    \begin{cases}
        I_0\frac{\frac{\pi}{2}-\arctan[r-(r_{ isco}-1)]}{\frac{\pi}{2}-\arctan[r_{ h}-(r_{ isco}-1)]}, &\quad r>r_{ h}\\
        0,&\quad r\leqslant r_{ h}
    \end{cases}.\label{diskprofile3}
\end{align}
The sketches of three accretion disk emission profiles are explicitly drawn in Fig.\ref{figprofile}.

\begin{figure}[htbp]
\centering
\subfigure[\, profile 1]
{\includegraphics[width=3.8cm]{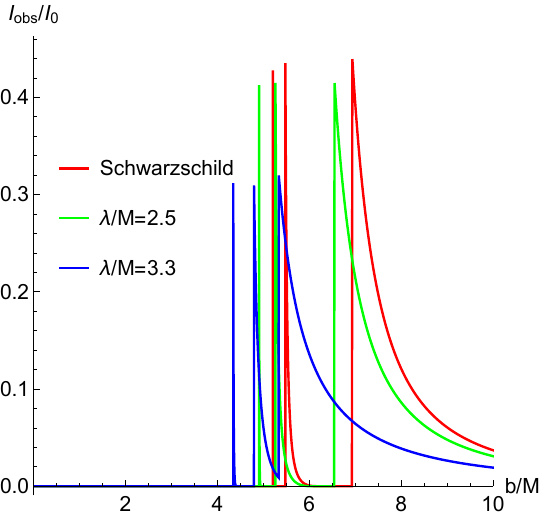} \label{}}\hspace{2mm}
\subfigure[\, Schwarzschild]
{\includegraphics[width=4.3cm]{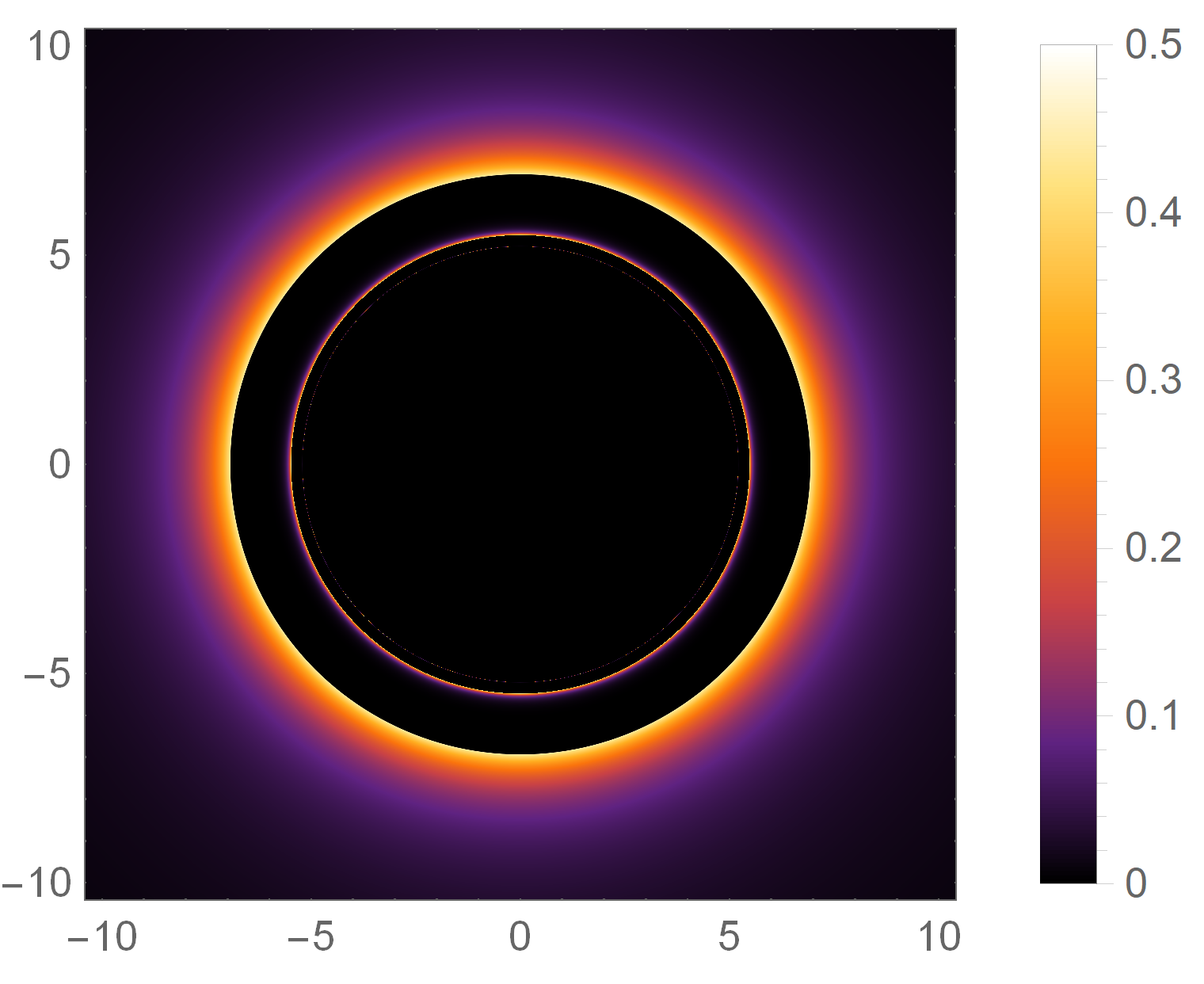}}\hspace{2mm}
\subfigure[\, $\lambda/M=2.5$]
{\includegraphics[width=4.3cm]{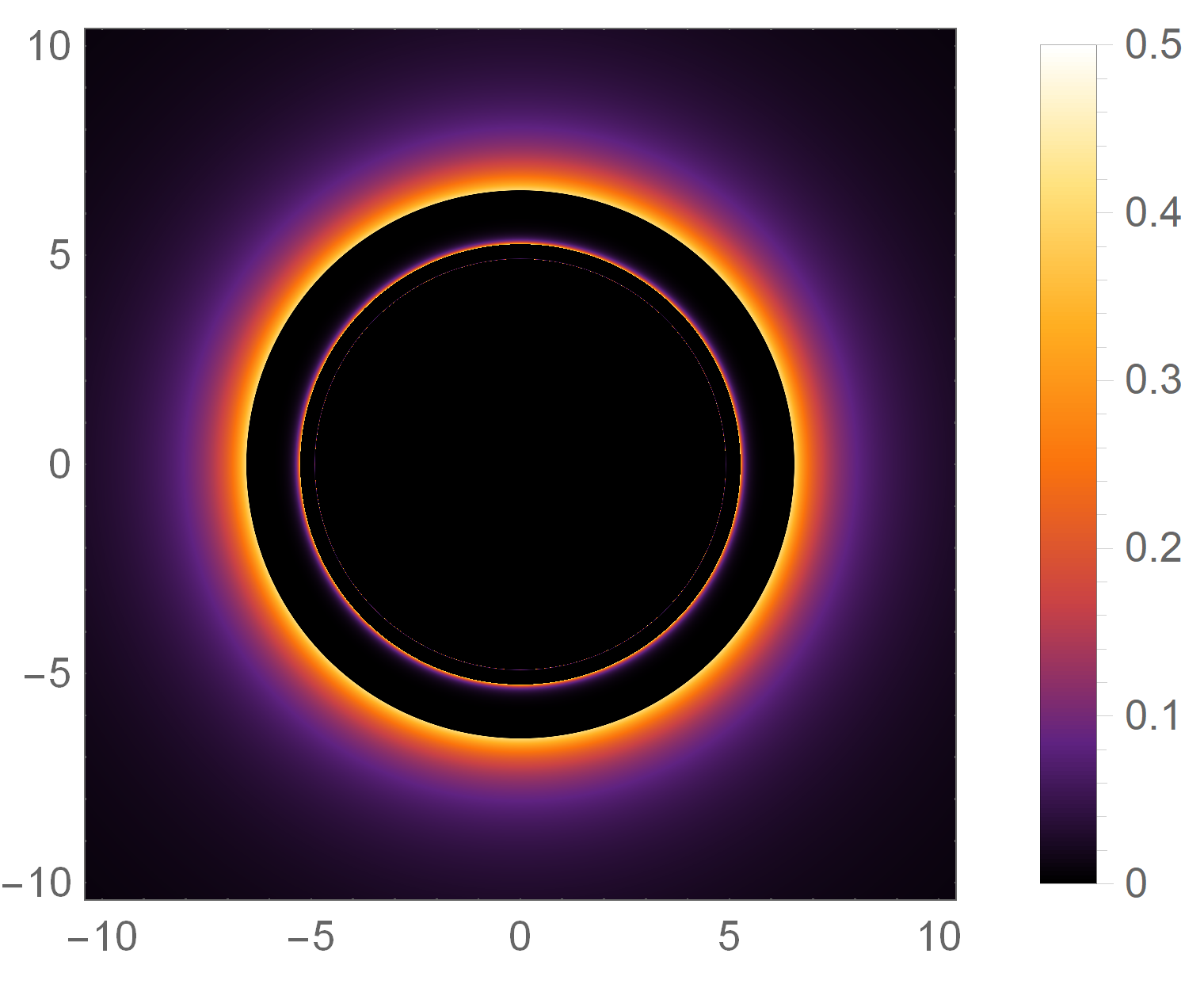}}\hspace{2mm}
\subfigure[\, $\lambda/M=3.3$]
{\includegraphics[width=4.3cm]{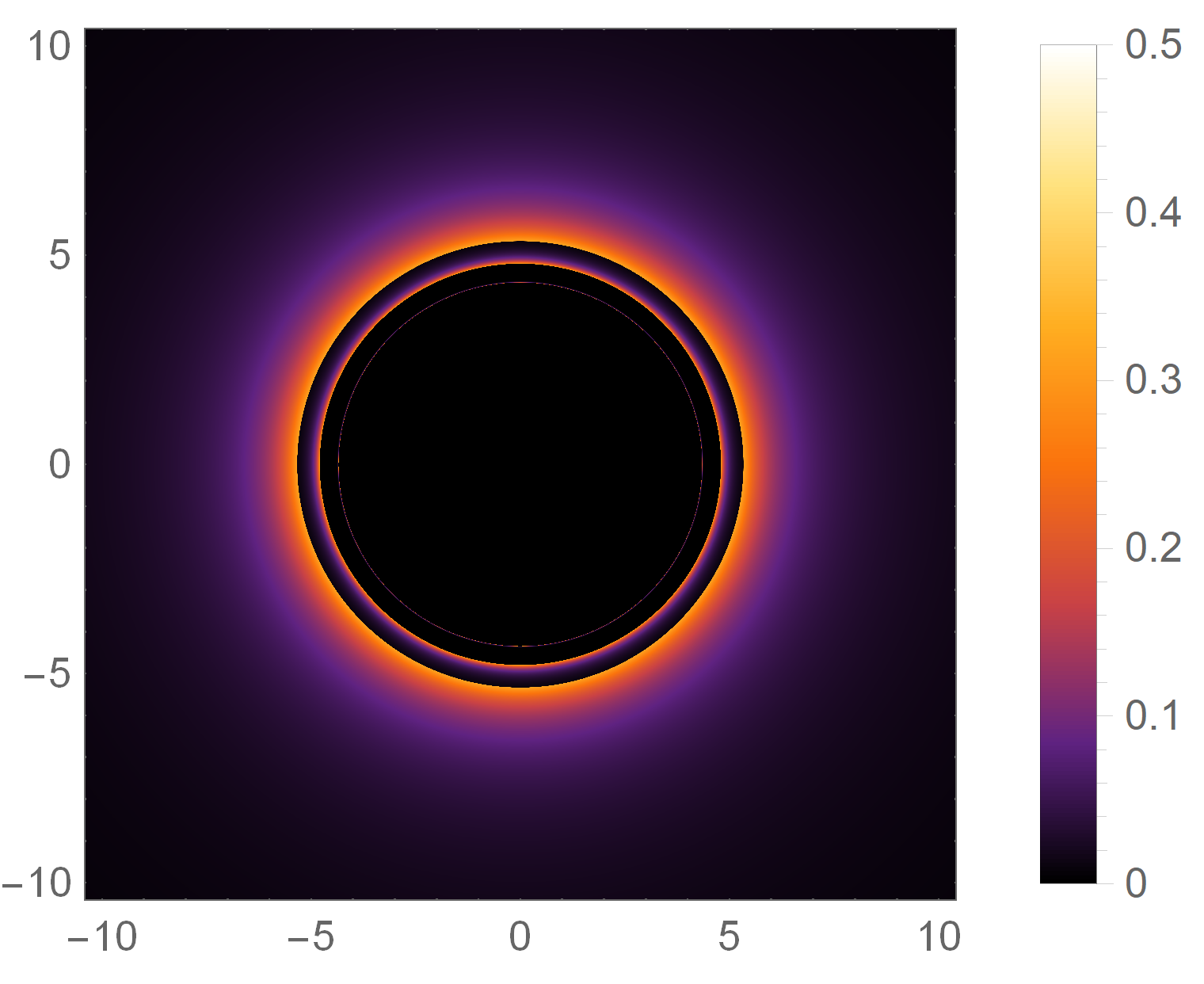}}\\
\subfigure[\, profile 2]
{\includegraphics[width=3.8cm]{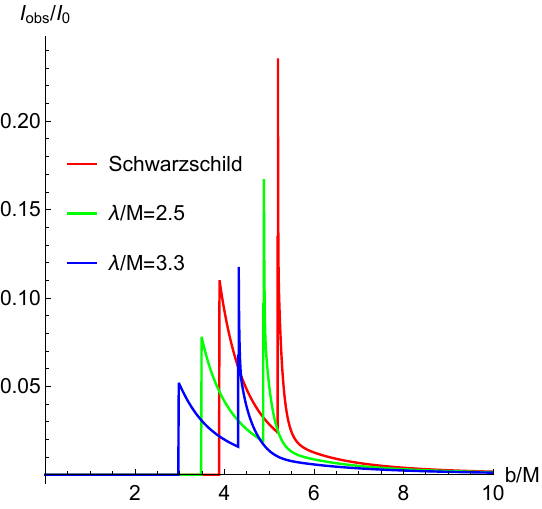} \label{}}\hspace{2mm}
\subfigure[\, Schwarzschild]
{\includegraphics[width=4.3cm]{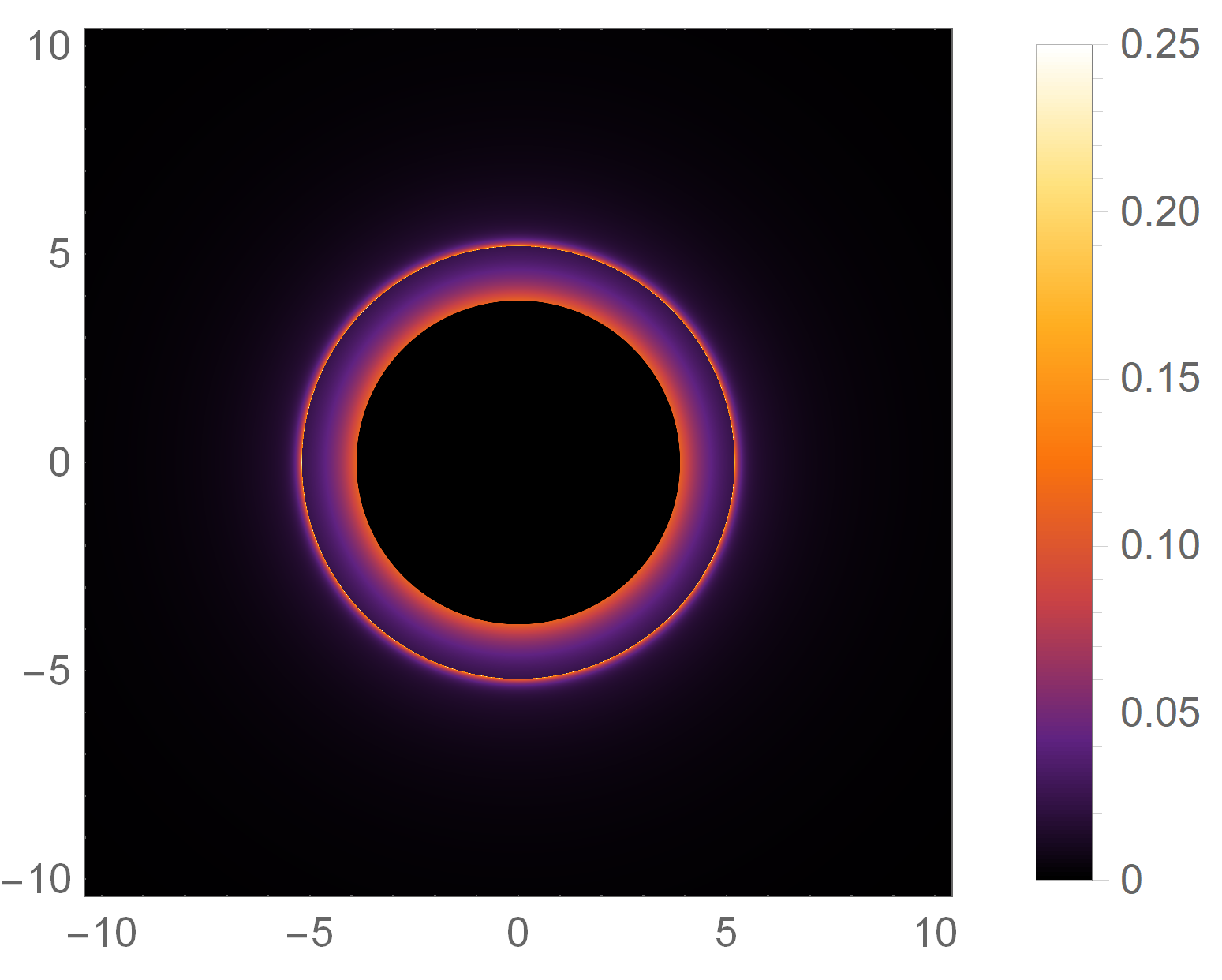}}\hspace{2mm}
\subfigure[\, $\lambda/M=2.5$]
{\includegraphics[width=4.3cm]{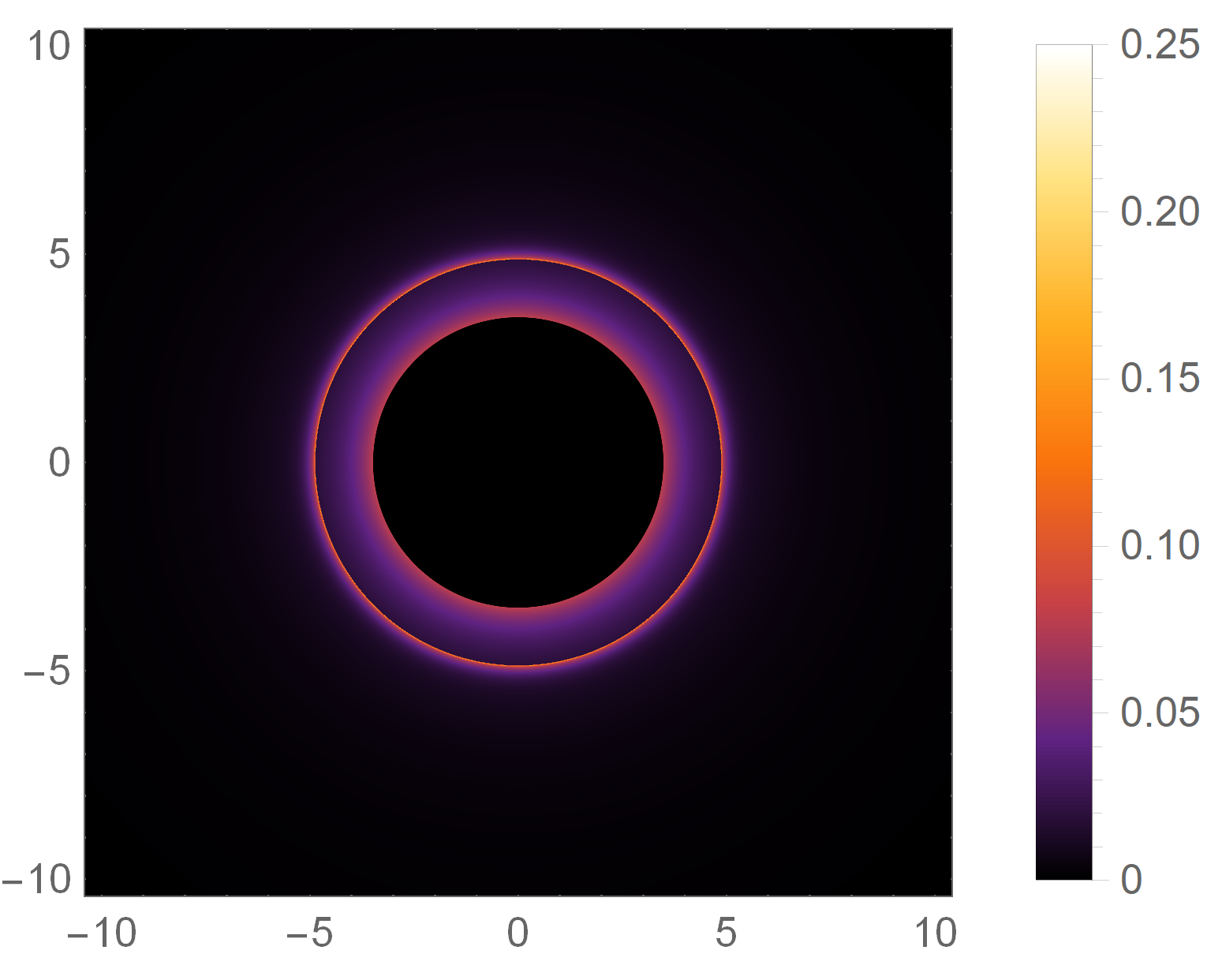}}\hspace{2mm}
\subfigure[\, $\lambda/M=3.3$]
{\includegraphics[width=4.3cm]{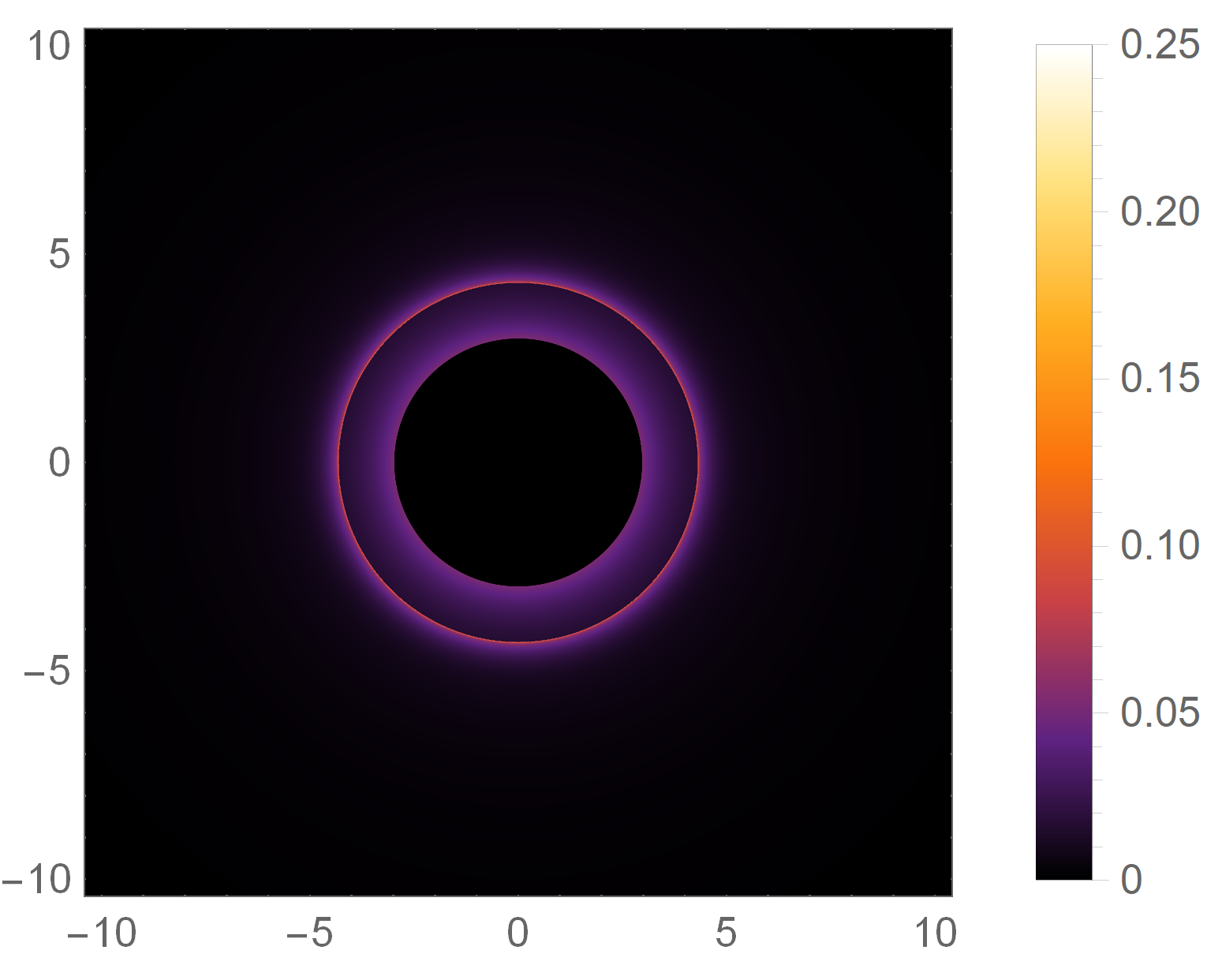}}\\
\subfigure[\, profile 3]
{\includegraphics[width=3.8cm]{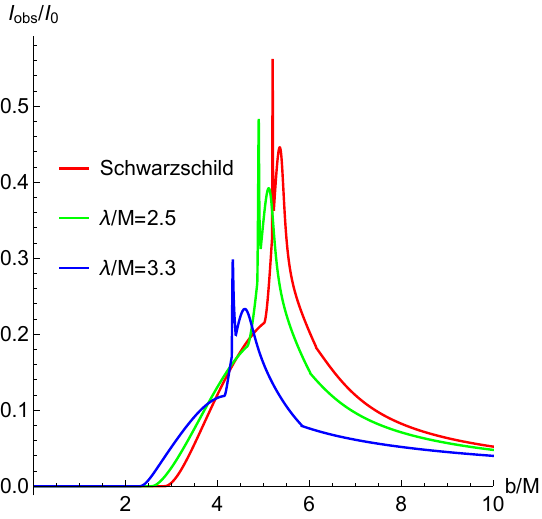} \label{}}\hspace{2mm}
\subfigure[\, Schwarzschild]
{\includegraphics[width=4.3cm]{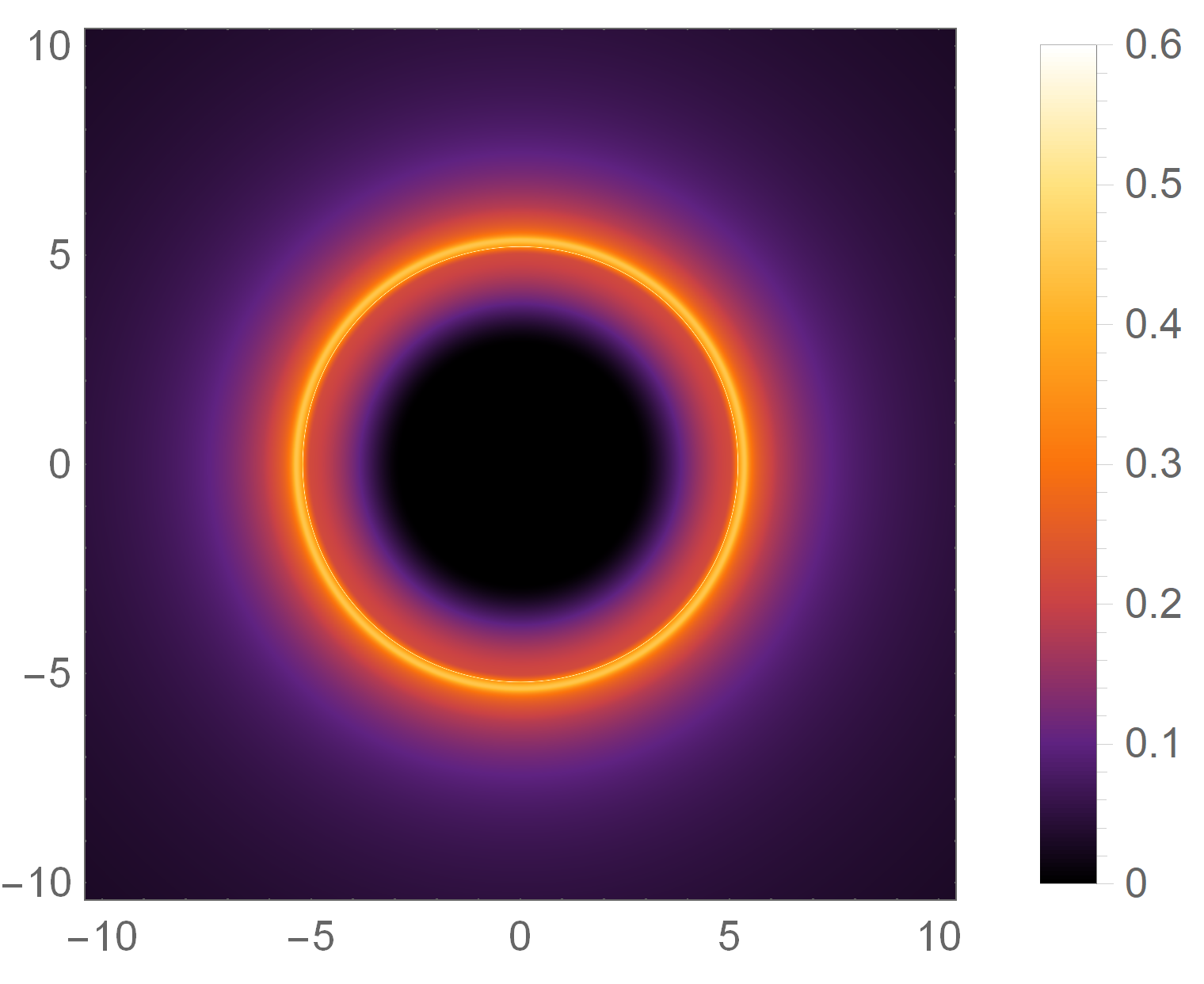}}\hspace{2mm}
\subfigure[\, $\lambda/M=2.5$]
{\includegraphics[width=4.3cm]{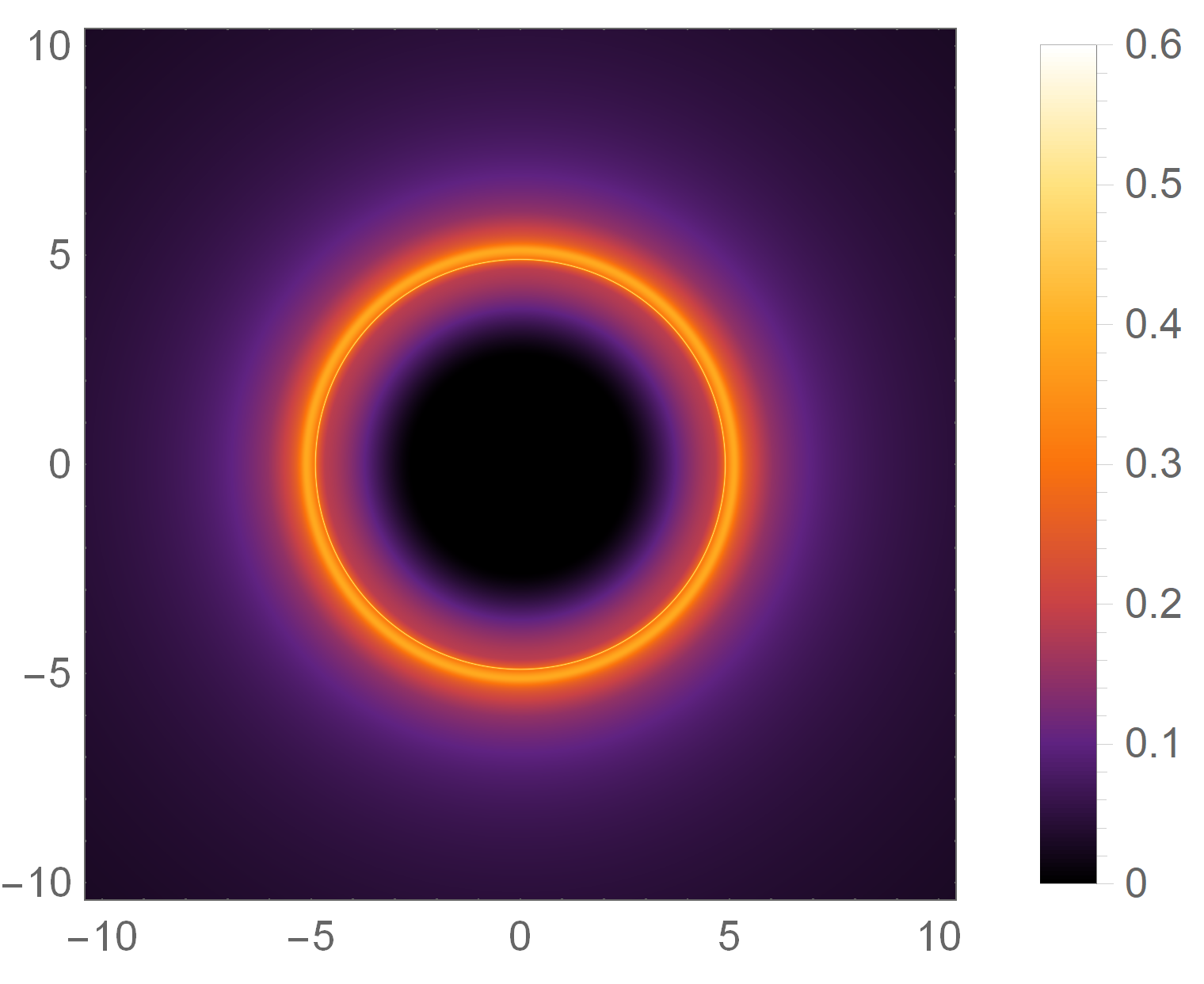}}\hspace{2mm}
\subfigure[\, $\lambda/M=3.3$]
{\includegraphics[width=4.4cm]{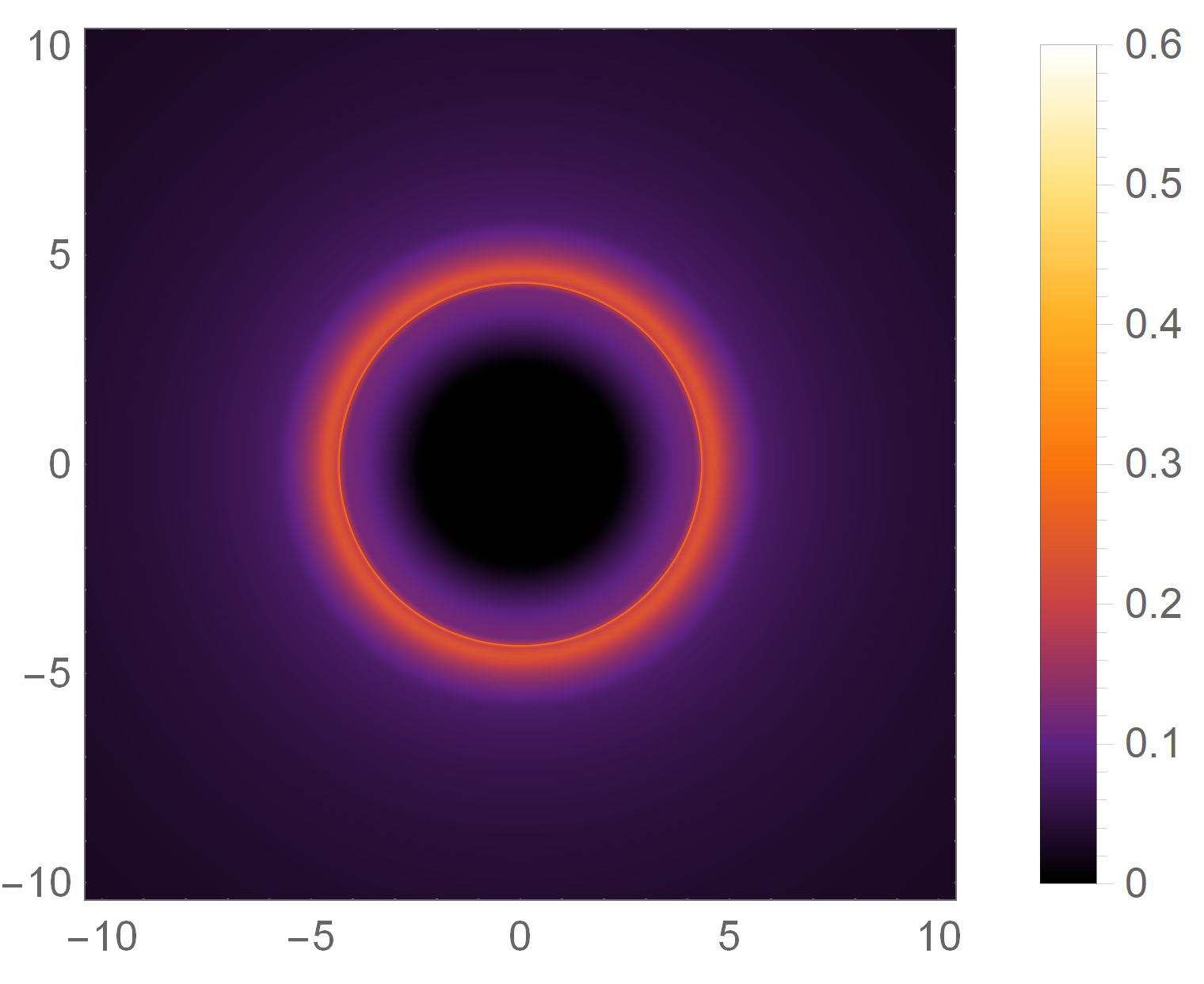}}\\
\caption{The total observed intensities (figures (a), (e), (i)) and observational appearances (other figures) of the emission profile 1 \eqref{diskprofile1}, profile 2 \eqref{diskprofile2} and profile 3 \eqref{diskprofile3} of the thin accretion disks for Schwarzschild and scalarized Gauss-Bonnet black holes respectively.}
\label{figdisk}
\end{figure}

Based on the given emission functions, we evaluate the total observed intensities for Schwarzschild and scalarized Gauss-Bonnet black holes, and then transfer them into the corresponding two-dimensional images, which are shown in Fig.\ref{figdisk}. For all cases, the results show that compared to Schwarzschild case, the scalarized black hole will show smaller shadow region and fainter brightness with the increasing of
coupling $\lambda$. Especially, for the first emission profile of accretion disk, we find that the middle peak in the observed intensities has a larger width than Schwarzschild case (in fact the innermost peak is also like this and just hard to see). This is because the demagnification factors of the second and third transfer functions are suppressed by $\lambda$ and the range of the corresponding emission becomes wider, as discussed in the above. Obviously, this leads the second ring of scalarized black hole is more clear than Schwarzschild case in the image. Similarly, for the second and third emission profiles, the peaks of scalarized black hole in the observed intensities are also wider than Schwarzschild cases by $\lambda$, indicating a wider but dimmed photon ring in the corresponding image.

\section{Rings and images of scalarized Gauss-Bonnet black hole illuminated by static spherical accretions}\label{spherical}

\begin{figure}[htbp]
\centering
\subfigure[\, observed intensity]
{\includegraphics[width=3.8cm]{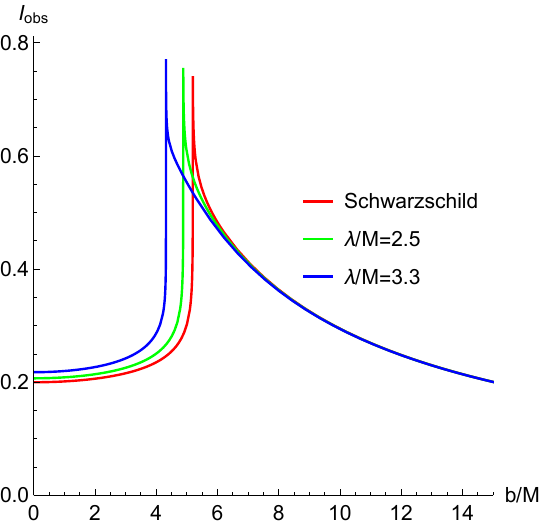}}\hspace{2mm}
\subfigure[\, $\lambda/M=0$]
{\includegraphics[width=4.3cm]{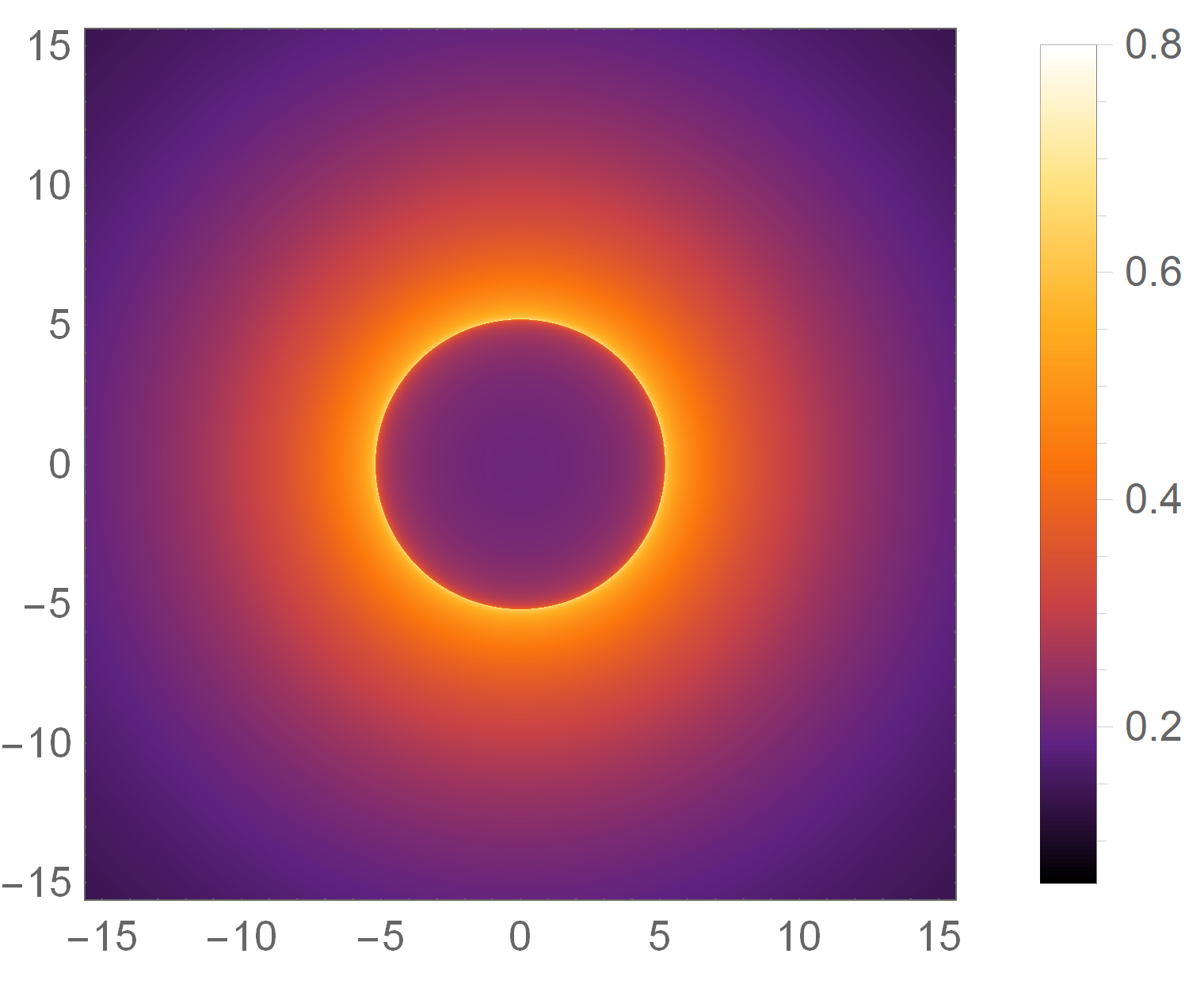}}
\subfigure[\, $\lambda/M=2.5$]
{\includegraphics[width=4.3cm]{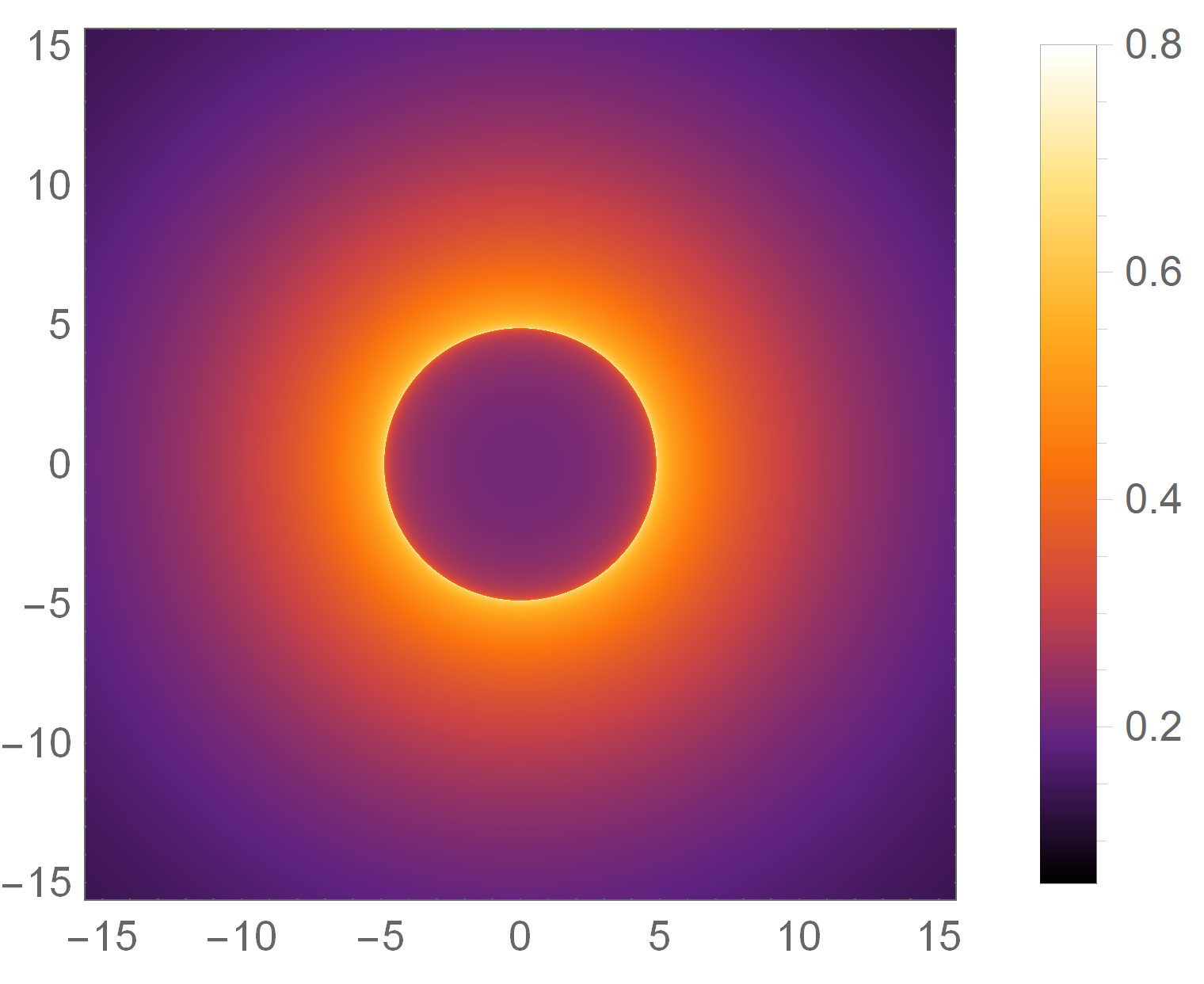}}\hspace{2mm}
\subfigure[\, $\lambda/M=3.3$]
{\includegraphics[width=4.3cm]{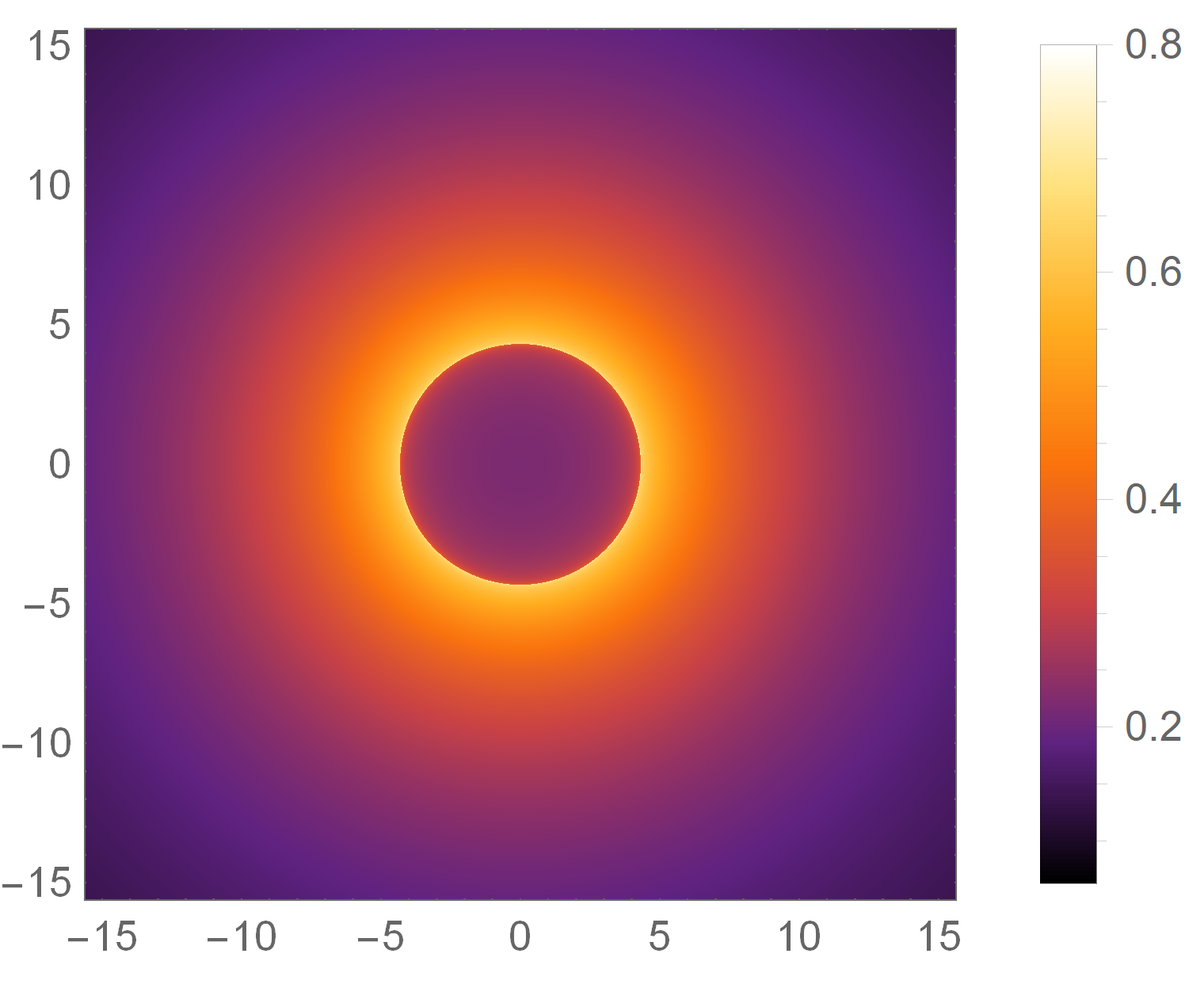}}
\caption{The total observed intensities (figure (a)) and observational appearances (other figures) of Schwarzschild and scalarized Gauss-Bonnet black holes surrounded by static spherical accretions, respectively.}
\label{figsph}
\end{figure}

In the above section, we consider the accretion disk surrounding the black hole which is formed by the materials in the Universe that are trapped by a black hole and rotate with a large angular momentum. Another situation is when the angular momentum is extremely small, the materials will flow radially to the black hole and form the spherically symmetric accretion \cite{Yuan:2014gma}. In this section, we will explore the images of scalarized black hole illuminated by the optically and geometrically
thin static accretion with spherical symmetry. Here, the observed
specific intensity $I(\nu_o)$ seen by an observer at infinity $r=\infty$  (measured in erg s$^{-1}$ cm$^{-2}$ str$^{-1}$ Hz$^{-1}$) is obtained by the as follows \cite{Bambi:2013nla}
\begin{equation}
I(\nu_o)=\int_{\gamma} g^3 j_e(\nu_e)dl_{prop},\label{eqIntensity}
\end{equation}
where $g=\nu_o/\nu_e=\sqrt{h(r)}$ is again the redshift factor \cite{Gan:2021pwu}. $\nu_o$ and $\nu_e$ are the observed and emitted photon frequency, respectively. $j_e(\nu_e)$ is the emissivity per unit volume in the rest frame and usually set $j(\nu_e)\propto\delta(\nu_r-\nu_e)/r^2$ where $\nu_r$ is rest-frame frequency of the emitter \cite{Bambi:2013nla}. $dl_{prop}$ is the infinitesimal proper length read as 
\begin{equation}
dl_{prop}=\sqrt{\frac{1}{f(r)}dr^2+r^2d\phi^2}=\sqrt{\frac{1}{f(r)}+r^2\left(\frac{d\phi}{dr}\right)^2}dr,
\end{equation}
where $d\phi/dr$ is given by Eq.\eqref{eqphoton}. So the formula Eq.\eqref{eqIntensity} indicates that the integration is along the photon path $\gamma$. Further, by integrating Eq.\eqref{eqIntensity} over the entire observed frequencies, we get the total observed intensity given by
\begin{equation}
I_{obs}=\int_{\nu_o} I(\nu_o)d\nu_o=\int_{\nu_e}\int_{\gamma}g^4j_e(\nu_e)dl_{prop}d\nu_e=\int_{\gamma}\frac{h(r)^{2}}{r^2}\sqrt{\frac{1}{f(r)}+r^2\left(\frac{d\phi}{dr}\right)^2}dr.\label{eqintensity}
\end{equation}
The total observed intensities and their corresponding images of scalarized black holes are shown in Fig.\ref{figsph}. For the observed intensities, we find that there exists always a peak which is located at $b=b_{ph}$. With the increase of the coupling $\lambda$, the maximum value of the peak increases which means the maximum brightness of scalarized black hole is larger than that of Schwarzschild case. Moreover, the size of black hole shadow decreases as the coupling $\lambda$ increases, which was also shown in Fig.\ref{figquantity2}(c). In the region inside shadow ($b<b_{ph}$), the observed intensity increases with $\lambda$. However in the region outside shadow ($b>b_{ph}$), it slightly decreases with $\lambda$. The observational appearances of scalarized black holes again confirm the above discussions.

\section{Conclusion and discussion}\label{conclusion}

EsGB theory, as one of the scalar-tensor theories, provides a theoretical framework that allows black hole with scalar hair. Specially, under certain conditions, there exist scalarized Gauss-Bonnet black hole solutions that are formed by spontaneous scalarization of the Schwarzschild black hole, which is induced by the curvature of the spacetime. The scalarized black hole solution and EsGB theory itself have been widely investigated from the theoretical and observational aspects. In this paper, we studied the optical features of the scalarized  Gauss-Bonnet black hole illuminated by various static thin accretions in EsGB theory to explore the observational differences of black hole with and without spontaneous scalarization.

Firstly, we gave a short review on the Gauss-Bonnet black hole with spontaneous scalarization in EsGB theory. Concretely, we briefly illustrated the reason for forming spontaneous scalarization of Schwarzschild black hole and provided the background solutions of scalarized black hole for some selected coupling parameters $\lambda$. Then we computed event horizon, the radius of photon sphere, critical impact parameter and ISCO influenced by $\lambda$. The results showed that these physical quantities decreases as $\lambda$ increases, which implied the smaller size of black hole shadow. Further, by using the currently observable data of black hole shadows from EHT, we gave the upper limit for the coupling. The observational  data gives the upper limits $\lambda/M=3.89$ by M87* black hole and $\lambda/M=3.31$ by Sgr A* black hole, respectively. Here we applied the shadow data with supermassive black holes, meaning that $M$ is quite larger than solar mass $M_\odot$, and constrain the coupling parameter for the coupling function \eqref{eq:coupling function}. It is noted that with the same coupling, the authors of \cite{Wong:2022wni} proposed to constrain spontaneous black hole scalarization with GW events GW190814 and GW151226 which refer to stellar black holes. Their simulations showed that GW190814 excludes a range value of $\lambda$ near $2M_1$ where $M_1\sim 23M_\odot$ is  the mass of the primary black hole in this GW event, while GW151226 cannot give any meaningful constraint. Since their black hole masses are much smaller than the supermassive we considered for shadow, so it is not straightforward to compare our constraints on the coupling parameter against theirs. It is optimistic that we have projects, such as Pulsar Timing Array (NANOGrav \cite{NANOGrav:2020bcs}, EPTA \cite{EPTA:2021crs}, PPTA \cite{Goncharov:2021oub}, IPTA \cite{Perera:2019sca}), space-based GW observatory (LISA \cite{LISA:2017pwj}, Taiji \cite{Ruan:2018tsw}, TianQin \cite{TianQin:2015yph}) and FAST \cite{Nan:2011um},  to detect nHz-MHz GWs released from supermassive black holes, which are hoped to be utilized to further constrain the spontaneous black hole scalarization in EsGB theory, and testify our constraints from supermassive black hole shadow.

Secondly, we considered the scalarized black holes illuminated by the optically and geometrically thin accretion disks. Comparing to Schwarzschild case, the lensed and photon ring emissions from the accretion disk correspond to
wider range of impact parameter for the scalarized black hole, and the demagnification factor in the second and third transfer functions are also suppressed. So the lensed and photon rings could be more easily to be observed. Then we considered three emission functions of accretion disks to compute the total observed intensities and images of scalarized black holes. For all models, we found that the scalarized black hole with larger coupling parameter $\lambda$ shows the smaller size of black hole shadow and weaker observed intensity but with the peak of larger width. Then we showed the scalarized black holes illuminated by the static spherical accretion. We found that a dark region indicated black hole shadow is surrounded by a bright photon ring. The shadow size for the scalarized black hole is smaller but the photon ring is more clear than that of Schwarzschild case.

In conclusion, we figured out the images of black holes with spontaneous scalarization illuminated by various static thin accretion disk and spherical accretion flow in EsGB theory. Compared to the black hole without spontaneous scalarization, our current results could disclose the significantly observational differences of black holes with spontaneous scalarization in EsGB theory. This preliminary study could provide a promising approach to test no-hair theorem by using black hole images. Here, we only consider the case of the finial state of scalarized black hole. So an interesting and meaningful direction is to investigate the image of black hole which is in the process of spontaneous scalarization. This will help us further understand the spontaneous scalarization of black hole and provide a potential method to test no-hair theorem from the observational perspective in the future.

\begin{acknowledgments}
This work is partly supported by the Natural Science Foundation of China under Grants Nos. 12375054 and 12222302, and  the Postgraduate Research $\&$ Practice Innovation Program of Jiangsu Province under Grants No. KYCX22$_{-}$3452.
\end{acknowledgments}


\bibliographystyle{utphys}
\bibliography{ref}

\end{document}